\documentclass[%
superscriptaddress,
 amsmath,amssymb,
 aps, physrev,
]{revtex4-2}
\usepackage{xcolor}
\usepackage{xspace}
\usepackage{graphicx}
\usepackage{dcolumn}
\usepackage{units}
\usepackage{bm}

\newcommand{\Xmax}{$X_{\rm max}$\xspace}

\begin{document}

\preprint{APS/123-QED}

\title{\textbf{A LOFAR-style reconstruction of cosmic-ray air showers with SKA-Low} 
}%

\author{A.~Corstanje}
\email{Contact author: a.corstanje@astro.ru.nl}
\affiliation{Vrije Universiteit Brussel, Astrophysical Institute, Pleinlaan 2, 1050 Brussels, Belgium}%
\affiliation{Department of Astrophysics/IMAPP, Radboud University Nijmegen, P.O. Box 9010, 6500 GL Nijmegen, The Netherlands}
\author{S.~Bouma}
\affiliation{Erlangen Centre for Astroparticle Physics, Friedrich-Alexander-Universität Erlangen-Nürnberg, 91058 Erlangen, Germany}
\author{S.~Buitink}
\affiliation{Vrije Universiteit Brussel, Astrophysical Institute, Pleinlaan 2, 1050 Brussels, Belgium}
\affiliation{Department of Astrophysics/IMAPP, Radboud University Nijmegen, P.O. Box 9010, 6500 GL Nijmegen, The Netherlands}
\author{M.~Desmet}
\affiliation{Vrije Universiteit Brussel, Astrophysical Institute, Pleinlaan 2, 1050 Brussels, Belgium}
\author{J.R.~H\"orandel}
\affiliation{Department of Astrophysics/IMAPP, Radboud University Nijmegen, P.O. Box 9010, 6500 GL Nijmegen, The Netherlands}
\author{T.~Huege}
\affiliation{Institut für Astroteilchenphysik, Karlsruhe Institute of Technology (KIT), P.O. Box 3640, 76021 Karlsruhe, Germany}
\affiliation{Vrije Universiteit Brussel, Astrophysical Institute, Pleinlaan 2, 1050 Brussels, Belgium}
\author{P.~Laub}
\affiliation{Erlangen Centre for Astroparticle Physics, Friedrich-Alexander-Universität Erlangen-Nürnberg, 91058 Erlangen, Germany}
\author{K.~Mulrey}
\affiliation{Department of Astrophysics/IMAPP, Radboud University Nijmegen, P.O. Box 9010, 6500 GL Nijmegen, The Netherlands}
\affiliation{Nikhef, Science Park Amsterdam, 1098 XG Amsterdam, The Netherlands}
\author{A.~Nelles}
\affiliation{Erlangen Centre for Astroparticle Physics, Friedrich-Alexander-Universität Erlangen-Nürnberg, 91058 Erlangen, Germany}
\affiliation{Deutsches Elektronen-Synchrotron DESY, Platanenallee 6, 15738 Zeuthen, Germany}
\author{O.~Scholten}
\affiliation{University of Groningen, Kapteyn Astronomical Institute, Groningen, 9747 AD, Netherlands}
\author{K.~Terveer}
\affiliation{Erlangen Centre for Astroparticle Physics, Friedrich-Alexander-Universität Erlangen-Nürnberg, 91058 Erlangen, Germany}
\author{S.~Thoudam}
\affiliation{Department of Physics, Khalifa University, P.O. Box 127788, Abu Dhabi, United Arab Emirates}
\author{K.~Watanabe}
\affiliation{Institut für Astroteilchenphysik, Karlsruhe Institute of Technology (KIT), P.O. Box 3640, 76021 Karlsruhe, Germany}

\date{\today}

\begin{abstract}
Cosmic-ray air shower detection with the low-frequency part of the Square Kilometre Array (SKA) radio telescope is envisioned to yield very high precision measurements of the particle composition of cosmic rays between $10^{16}$ and $\unit[10^{18}]{eV}$. This is made possible by the extreme antenna density of the core of SKA-Low, surpassing the current most dense radio air shower observatory LOFAR by over an order of magnitude. In order to make these measurements, the technical implementation of this observation mode and the development of reconstruction methods have to happen hand-in-hand. As a first lower limit of what is obtainable, we apply the current most precise reconstruction methods as used at LOFAR to a first complete simulation of air shower signals for the SKA-Low array. We describe this simulation setup and discuss the obtainable accuracy and resolution. A special focus is put on effects of the dynamic range of the system, beamforming methods to lower the energy threshold, as well as the limits to the mass composition accuracy given by statistical and systematic uncertainties. 
\end{abstract}

\maketitle

\section{\label{sec:intro}Introduction}

Extensive air showers induced by cosmic rays have been measured with many techniques to address the long-standing question about their origin. Measuring the radio emission of air showers has reached competitive performance in the past decade by showing the high accuracy that can be achieved in the reconstruction of the energy of the shower \cite{PierreAuger:2015hbf,PierreAuger:2016vya,Auger_energy:2024} and the depth of shower maximum \cite{Buitink:2014eqa,Auger_radio:2024}. As predicted in the 1960s, the radio emission stems from the charge distributions in the electromagnetic cascade in the air shower \cite{Glaser:2016qso,1966RSPSA.289..206K}, and is therefore a calorimetric measurement that also traces the shower development. 
	
Radio detection of air showers has become a reliable and accurate method for the reconstruction of the energy, direction, and mass composition of cosmic rays. Several antenna arrays have been constructed specifically for this purpose, either stand-alone (e.g.\ CODALEMA \cite{Ardouin:2005}) or as a complementary component to existing cosmic-ray observatories (e.g.\ LOPES \cite{Falcke:2005} and the radio array of the Pierre Auger Observatory \cite{Huege:2019}). Alternatively, radio observations of air showers can be performed with general-purpose radio telescopes, provided that these telescopes consist of omnidirectional antennas and offer buffer and trigger capabilities. 
This allows to record cosmic-ray signals completely in parallel to ongoing astronomical observations, an otherwise unusual observation mode.

LOFAR \cite{vanHaarlem:2013} was one of the first radio telescopes that has such a design and has played an important role in development of cosmic-ray detection and reconstruction techniques, and the validation of the theory of the air shower radiation mechanisms \cite{Corstanje:2014waa,Schellart:2014oaa,Nelles:2014dja,Scholten:2016gmj}. Whereas dedicated cosmic-ray radio arrays are usually designed to optimize the total instrumented area, radio telescopes require a different design where antennas are placed in positions that create a large variety in baselines to optimize the UV coverage in interferometric observations. As a result, the antenna density of radio telescopes is much larger than that of dedicated cosmic-ray arrays, but concentrated in a smaller total area, which constrains the maximum cosmic-ray energy that can be reached. LOFAR reaches a precision of $\unit[9]{\%}$ ($\pm \unit[14]{\%}$ syst.) in energy and $\unit[19]{g/cm^2}$ ($\pm \unit[9]{g/cm^2}$ syst.) in the shower maximum, \Xmax, in an energy range of $10^{16.7}$ to $\unit[10^{18.3}]{eV}$ \cite{Corstanje:2021kik,Buitink:2016nkf}. In the near future, the Square Kilometre Array (SKA) offers the prospect of obtaining an even better resolution, as demonstrated in this paper.

Construction for the SKA has started in Australia and South-Africa. The low-frequency component in the Australian outback will consist of nearly sixty thousand log-periodic dipole antennas on an area of roughly one square kilometer. Individual air showers will be simultaneously observed by thousands of antennas, which is an order of magnitude higher than what is obtained at LOFAR. Moreover, the SKA will sample the radiation footprint on the ground much more homogeneously than LOFAR. The frequency bandwidth of the SKA-Low array (50-350 MHz) is broader and higher than LOFAR (30-80 MHz), which has various benefits. For example, the shape of the frequency spectrum carries valuable information on the shower geometry \cite{Welling:2019scz}. In addition, the Galactic background noise is low at higher frequencies \cite{Razavi:2011}, offering an excellent signal-to-noise ratio for antennas close to the Cherenkov ring (where a high-frequency component is strongly present) \cite{Welling:2019scz}. While the highest energy that can be reached is still limited by the surface area (to roughly $\unit[10^{18}]{eV}$), the increased bandwidth and high antenna density make it possible to reach much lower energies than most other radio arrays. As shown in this analysis, by employing beamforming techniques, air shower reconstruction is possible down to at least $\unit[10^{16}]{eV}$, and possibly even lower. 

Thus, the SKA cosmic-ray energy range spans the region roughly between the knee and ankle in the spectrum, where a transition is expected between a Galactic and extragalactic flux \cite{ParticleDataGroup:2024cfk}. In addition, there is a case for the existence of a secondary Galactic component \cite{Thoudam:2016}, for which possible scenarios are re-acceleration on the Galactic termination shock, or shockwave acceleration in supernovas expanding into the strongly magnetized winds surrounding Wolf-Rayet stars. 

To disentangle these contributions, it is necessary to perform accurate measurements of the mass composition. While important progress has been made by either studying the electron/muon ratio (e.g. KASCADE-Grande \cite{Brancus:2005ht} and IceTop \cite{Abbasi:2021RF}) or the depth of shower maximum with fluorescence or Cherenkov light (e.g. Auger \cite{Auger_depth:2014}, TALE \cite{TALE:2020}, Yakutsk \cite{Knurenko:2015}), many uncertainties remain. The extremely dense antenna array of the SKA will provide an unprecedented precision in this energy range, potentially leading to a better understanding of the highest-energy Galactic sources and the onset of the extragalactic component. 

Systematic uncertainties introduced by the modeling of high-energy hadronic interactions, beyond the energies accessible by particle colliders like LHC, are another obstacle to accurate measurements.
State-of-the-art models like QGSJET \cite{Ostapchenko:2013}, EPOS-LHC \cite{EPOSLHC:2013} and SIBYLL \cite{Sibyll:2020} predict different shower developments which results in a different mass interpretation of the measurements. The high-resolution observations of SKA-Low can be used to constrain the shape of the longitudinal development of air showers and thereby provide a new way of putting these models to the test \cite{Buitink:2023reh}. To evaluate such possibilities and make use of all the information provided by the high-density, large-bandwidth observations, new reconstruction techniques are under development that leverage the unique capabilities of the SKA. 

In this paper, we apply the most precise, existing reconstruction method as used at LOFAR to air shower simulations for SKA. Likewise, it aims at accurately measuring the depth of shower maximum, \Xmax as the main quantity sensitive to the primary particle mass.
We study the expected accuracy for the measurement of \Xmax, as well as the expected primary energy range that can be probed with SKA using established techniques. 
We also discuss implications of SKA design choices for air shower detection with SKA. In particular, we consider the available dynamic range in the analog-to-digital converters, and the number of measured air showers and their data volumes that are required to make a major improvement on LOFAR results and to make proper use of the capabilities of this next-level instrument.

New simulation techniques are being developed that allow to produce radio traces for a given longitudinal shower development \cite{Desmet:2025}, much faster than by doing a full microscopic simulation. This will reduce computation time for the anticipated large number of measured showers.
Also, newer analysis techniques being developed are expected to improve on the results presented here, for instance by measuring the longitudinal shower evolution in more detail \cite{Corstanje:2023uyg}; therefore the present results can be regarded as a baseline of the performance of SKA-Low.

\begin{figure}
	\centering
	\includegraphics[width=0.7\textwidth]{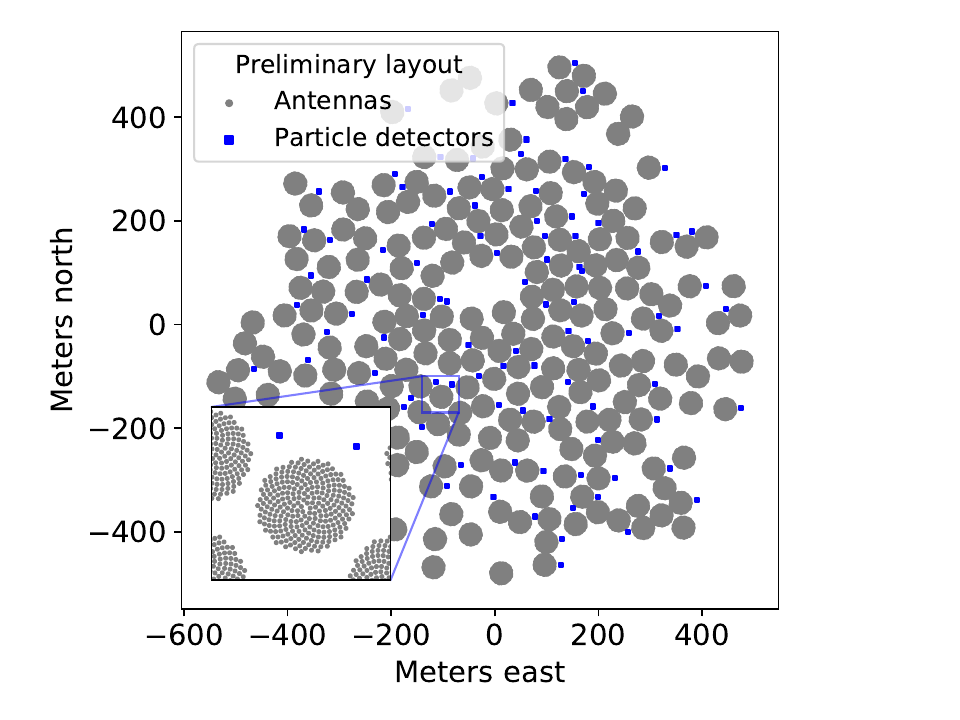}
	\caption{The antenna layout of the SKA-Low inner core region (layout within stations is not definitive), with 100 particle detectors of $\unit[1]{m^2}$ (points not to scale) placed quasi-randomly adjacent to antenna stations, at a minimum distance of $\unit[10]{m}$. The exact positions do not matter much for air shower detection and will be decided based on logistical and operational needs of SKA.}
	\label{fig:ska_layout_with_particle_detectors}
\end{figure}

\section{Reconstructing the shower maximum}
\label{sec:method}
Exploring the capabilities of SKA-Low requires both a complete simulation of the measurement, as well as an adaptation of the reconstruction process used at LOFAR \cite{Buitink:2014eqa,Corstanje:2021kik} to the SKA simulations. We first elaborate on the simulation setup and the specifics of SKA that had to be taken into account. We then describe the reconstruction algorithm adapted from LOFAR to obtain \Xmax and discuss simple improvements possible through the antenna density of SKA such as beamforming of groups of antennas to increase signal-to-noise ratios (SNR) when measuring lower-energy cosmic rays.

We would like to emphasize that actual observing with SKA will require dedicated methods that go beyond what is currently used within the community, which will improve the reconstruction. In particular, the present method was designed for a more narrow bandwidth (30 to $\unit[80]{MHz}$) and for measuring the shower maximum \Xmax only. Using additional spectral information in a 50 to $\unit[350]{MHz}$ bandwidth, and measuring more parameters of the longitudinal distribution of particles such as explored in \cite{Buitink:2023reh,Corstanje:2023uyg} are expected to bring further improvements in both \Xmax accuracy and mass composition analysis.
This is why the present description should be seen as lower limit of what is possible. 

\subsection{Simulation setup for SKA-Low}
The simulations are based on CORSIKA \cite{Heck:1998vt}, using in particular CoREAS to obtain the radio signals \cite{Huege:2013vt}. Simulations were done specifically for the site parameters of SKA-Low (altitude $\unit[378]{m}$, magnetic field horizontal and vertical component $\unit[27.60]{\mu T}$ and $\unit[-48.27]{\mu T}$ respectively) on a generic `star-shaped' radial grid \cite{Buitink:2014eqa} of 208 antennas. Simulations were performed at a single primary energy of $\unit[10^{17}]{eV}$. As it has been shown by various theoretical and experiment studies, e.g.~\cite{PierreAuger:2015hbf,Glaser:2016qso,Gottowik:2017wio}, the signal amplitude scales linearly with the shower energy. Thus, the electric field levels of the radio signals are scaled according to the desired primary energy level. The \Xmax distributions also change as a function of energy, which needs to be taken into account for a composition measurement. However, using the same distributions gives sufficient coverage of the \Xmax range for a resolution study, especially as we use relatively large ensembles of 140 showers as described below. 

To produce electric field traces on the nearly 60,000 antennas in the SKA-Low core region, we make use of the full signal interpolation method in \cite{2023JInst..18P9005C}, that was found to produce results with an accuracy suitable for high-precision work. This saves over two orders of magnitude of computing time compared to direct simulation of all antennas.
We use an implementation of the official model of the SKA-Low antennas (SKALA4 \cite{NuRadioMC_code,Glaser:2019cws}) to convert the electric field signals to measured voltages at the antennas.
We apply an additional unit-gain filter to de-disperse the pulses, i.e.~compensating for the phase delays introduced by the antenna model. As simulated cosmic-ray pulses have a nearly flat phase spectrum, this optimizes the signal-to-noise ratio of the pulses, and makes them narrower which helps estimate their energy fluence in a short time window. Since there is currently no model for the full system response, this approach is considered a good compromise, especially as phase spectra of many strong pulses can also be measured in practice once real data is available.

The SKA system is designed to be sky-noise dominated. Realistic levels of noise from the Galaxy (model GSM2016 \cite{Zheng:2016lul}) were calculated using the NuRadio software framework \cite{Glaser:2019cws} and added to the voltage signal traces. We have taken a noise level around the median level found in a sidereal day. In accordance with conservative design estimates, instrumental noise has been modeled as flat-spectrum noise at $\unit[30]{\%}$ of the integrated Galactic noise power as an approximation of its magnitude. When more characteristics of the signal chain are known, in particular active components like the low-noise amplifier (LNA) can give rise to noise contributions with a non-flat spectrum. For the present analysis these are not yet essential to model.

\begin{figure}
	\centering
\includegraphics[width=0.49\textwidth]{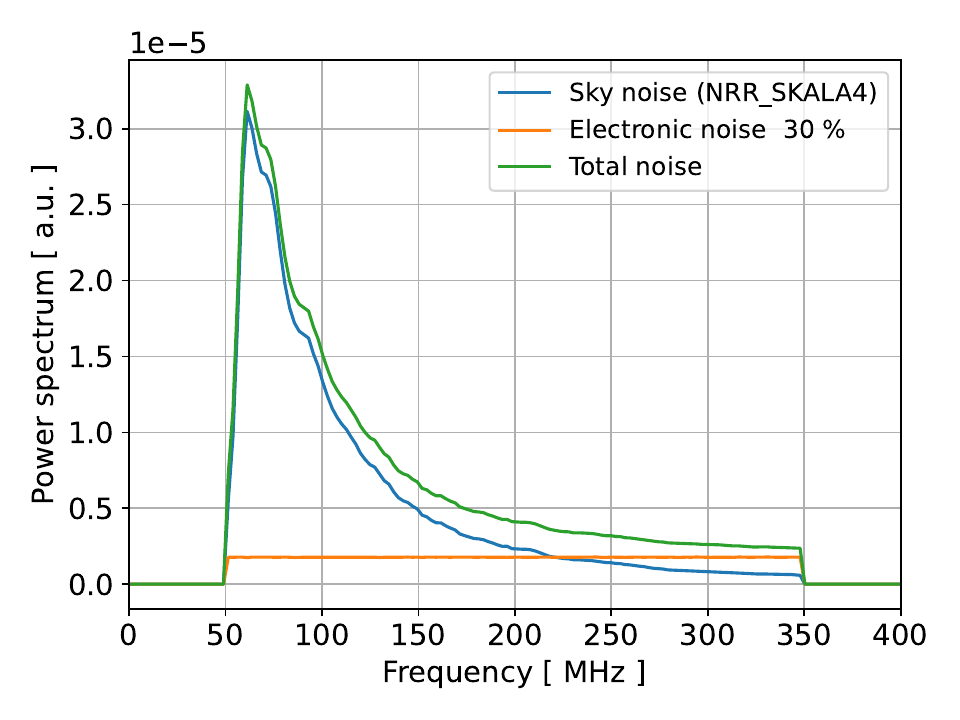}
	\includegraphics[width=0.49\textwidth]{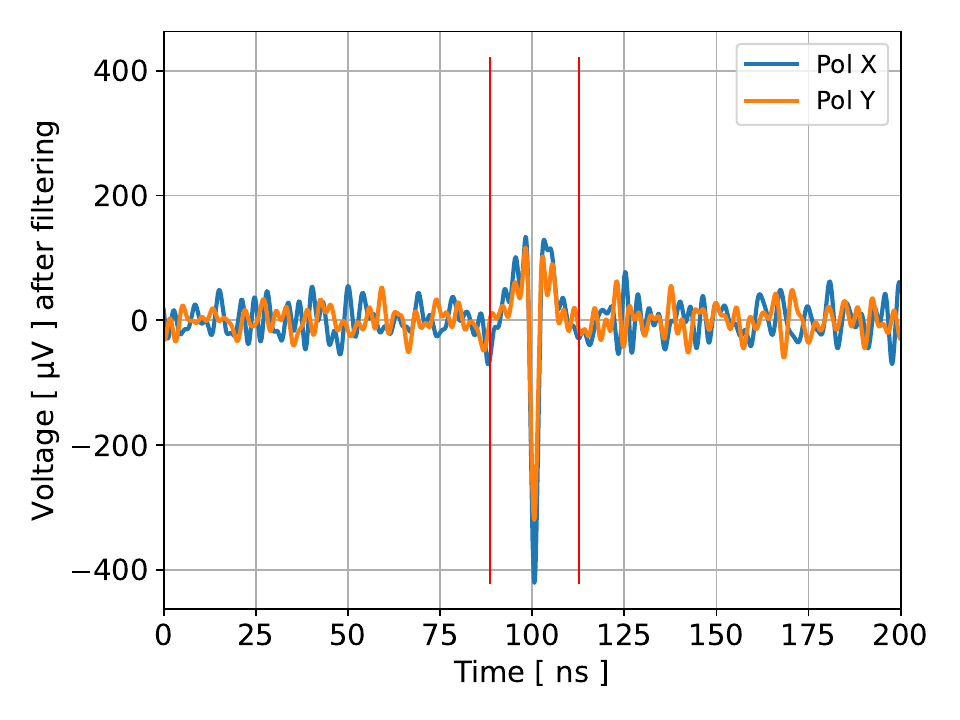}
	\caption{Left: an average power spectrum of the simulated noise at an antenna receiver. The contributions are shown from the Galactic background as well as flat-spectrum electronic noise at an assumed level of $\unit[30]{\%}$ of the total Galactic noise power. Right: an example of an air shower voltage trace with a noise background, after filtering. Vertical lines indicate the time window used to measure the fluence. The frequency band starting at $\unit[50]{MHz}$ (well away from zero) gives rise to the wider features around the sharp central pulse.}
	\label{fig:noise_power_spectrum}
\end{figure}

The average noise power spectrum in a voltage trace is shown in Fig.~\ref{fig:noise_power_spectrum}, where we also show a time-domain trace of an example cosmic-ray radio pulse from a $\unit[10^{17}]{eV}$ shower with noise added.
The power spectrum has a considerable falloff which arises from the sky temperature of the Galactic background, which in our frequency range has a power-law spectrum with a spectral index near $-2.55$:
\begin{equation}
T(\nu) \propto (\nu / \nu_0)^{-2.55}.
\end{equation} 
Unlike at LOFAR, where antennas have a narrower bandwidth and a notable resonance near $\unit[58]{MHz}$, the spectrum falloff remains appreciable when measured after the antenna response.

Therefore, to weight to the signal content across the frequency band by the SNR, we apply {\it noise whitening} to the simulated signals, dividing out the noise power spectrum in the frequency domain. This means that after this filter, the total noise spectrum becomes flat on average.
This would also optimize the signal-to-noise ratio for an ideal (flat-spectrum) bandwidth-limited pulse. The strongest signal pulses in the radio footprint, i.e.~those that remain visible for the weakest detectable showers, have a nearly flat spectrum. Thus, this additional filter helps to improve the low-energy detection limit. Also, noise with a flat spectrum has lower correlations in its time series (if there were no bandwidth limit it would be uncorrelated), which is helpful for fluence measurements in a short time window.

This procedure only depends on the noise spectrum which is readily measurable in each dataset, without dependence on pulse simulation results.
The noise spectrum will vary somewhat from one dataset to another, as the sky noise RMS level varies by about $\pm \unit[25]{\%}$ around the median level during a day. As the electronic noise is expected to be mostly constant, the (combined) shape will change, although not by much (see Fig.~\ref{fig:noise_power_spectrum}).
If a single, constant noise whitening filter is desired, not depending on time of day, a time-averaged compromise can be found at a minor expense of SNR performance (likely at a level below $\unit[10]{\%}$ which is negligible for this study).

\begin{figure}
	\centering
\includegraphics[trim={1.0cm 0.5cm 1.0cm 0.5cm},clip,width=0.49\textwidth]{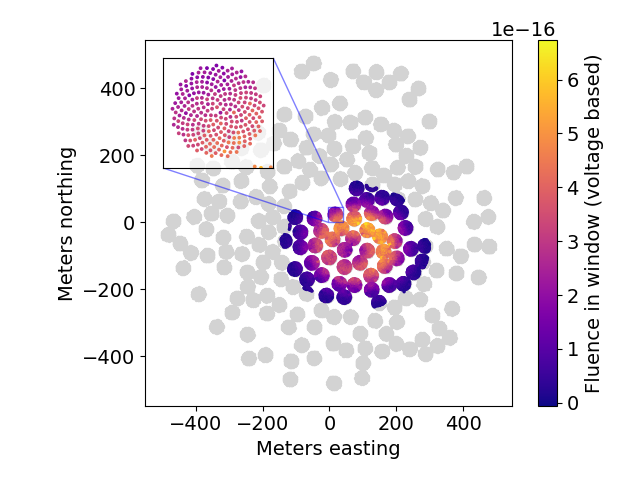}
	\includegraphics[trim={1.0cm 0.5cm 1cm 0.5cm},clip,width=0.49\textwidth]{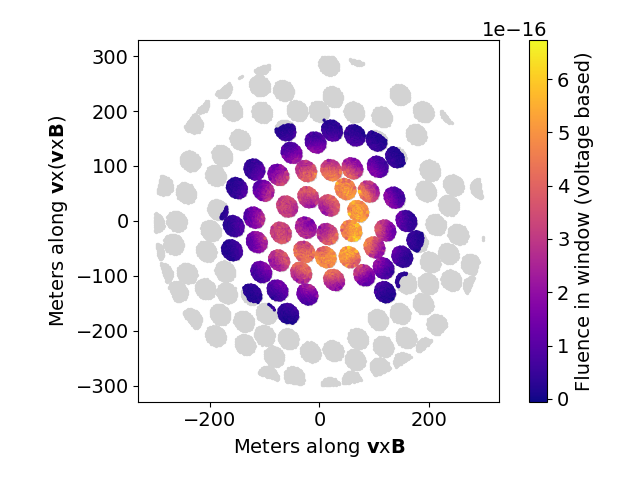}
	\caption{An example of how a measured radio fluence footprint of a cosmic-ray air shower would look like, for antennas that meet a $\unit[5]{\sigma}$ amplitude criterion. Left panel: antenna stations, each comprising 256 antennas, are shown in ground plane coordinates. The inset shows the details within a station. Right panel: the footprint projected onto the shower plane, perpendicular to the incoming direction of the cosmic particle $\mathbf{v}$. Here, antennas are included up to $\unit[300]{m}$ from the shower core for performance reasons, omitting antennas with zero signal.}
	\label{fig:footprint_example}
\end{figure}

\subsection{Fluence measurements}
The shower reconstruction is based on fluence measurements at voltage level, i.e., after the antenna response and after the two post-filters described above. 
In a practical setting, it will also have passed the other components of the signal chain such as the amplifier, which are essentially also described by filters.

The fluence is defined as a time integration of $V^2$, i.e.~voltage-squared, in a short time window.
This window is centered on the peak of the pulse as obtained from a Hilbert envelope of the trace, see Fig.~\ref{fig:noise_power_spectrum}, right panel. It can be symmetric around the peak, as after the de-dispersion filter, the pulse is symmetric, unlike the (causally filtered) pulse as it is measured directly.
The average noise background level is subtracted, to obtain an unbiased estimate of the fluence (at least for "ideal" filtered Gaussian noise). 
We note that the treatment of small fluence signals and their bias and uncertainty is the subject of discussion, see e.g.~\cite{Martinelli:2024bzg}. This is at this point not yet perfectly resolved, as is seen e.g.~from the need for an amplitude threshold as described below. However, as the results in the next sections show, the method used here was found to perform adequately for the purpose of this study, which is setting a baseline result.
Hence, the estimated fluence $f_{\mathrm{est}}$ from a time trace $x(t)$ is defined as
\begin{equation}\label{eq:fluence_measurement}
    f_{\mathrm{est}} = \sum_t^{N_s} x(t)^2 - N_s \sigma_n^2,
\end{equation}
where $N_s$ is the number of samples in the window, and $\sigma_n$ is the standard deviation of the noise.

The length of the window is somewhat arbitrary, and is chosen from a trade-off between capturing as much as possible from the energy fluence, and picking up as little noise as possible. A length of $\unit[24]{ns}$ was found to work reliably. This number has not been explicitly optimized, and results are not sensitive to values e.g.~between 20 and $\unit[30]{ns}$. At this length, 90 to $\unit[97]{\%}$ of the energy fluence is contained in the window, with the highest fraction for strong pulses near the Cherenkov cone such as the example shown in Fig.~\ref{fig:noise_power_spectrum}. This satisfies the requirement for a good signal/noise trade-off. As the same procedures are applied to data as to simulations, measuring somewhat less than $\unit[100]{\%}$ of the fluence is not a problem for fitting, as long as the antenna and electronics response have been modeled and/or measured accurately.

We position the fluence integration window at the (noiseless) signal peak time, anticipating an extrapolation of the measured wavefront to those antennas with insufficient signal strength to accurately determine the peak time. Remaining uncertainties on the peak time are expected to be within about $\pm \unit[3]{ns}$, i.e.~much smaller than the signal integration window of $\unit[24]{ns}$, which is a small fraction of the anticipated $\unit[50]{\mu s}$ that would be stored as data following a trigger.

The uncertainty on the measured fluence is estimated separately using a Monte Carlo analysis, adding 1000 random noise realizations to the simulated traces, and comparing the measured fluence to the known real fluence. 
For uncorrelated Gaussian noise, the variance of a fluence measurement is given by
\begin{equation}\label{eq:fluence_uncertainty}
\sigma_f^2 = 4\sigma_n^2\,f + 2 N_s \sigma_n^4,
\end{equation}
which is a linear function in the true fluence $f$.
For filtered Gaussian noise, as used here, the variance is still well described by such a linear function, but its coefficients are different, as in 
\begin{equation}\label{eq:fluence_uncertainty_practical}
\sigma_f^2 = a\,f + b,
\end{equation}
and we fit the coeffients $a$ and $b$ using our noise Monte Carlo results.
We assign uncertainties to the fluence estimates based on this function; however as seen from Eq.~\ref{eq:fluence_measurement}, the subtraction of the average noise background sometimes leads to negative fluence estimates when the true fluence is weak or zero. This is inevitable for an unbiased estimator of a non-negative quantity.
It happens only rarely when a minimum amplitude criterion is used (see below).
Negative fluence estimates are used as they are, i.e.~not capped at zero, and they are assigned the uncertainty at zero fluence.

We have found that the weakest measurements, at a fluence near zero, may introduce a bias towards smaller or larger footprints. This might be caused by an underestimate of their uncertainty, or by a (very small) bias in the fluence measurement near zero values. We apply a scale factor $5$ to the uncertainty at zero fluence by enlarging the constant term in Eq.~\ref{eq:fluence_uncertainty}, thus reducing the weight of these measurements in a chi-squared fit. Uncertainties at appreciable fluence levels are hardly affected by this. Consequently, the reduced chi-squared values for an optimal fit are expected to be below unity.

We apply an amplitude threshold of $5\,\sigma$ that the Hilbert envelope of the signal in one polarization must reach for an antenna to be included in the fluence fits.
This ensures that at least some antennas have a strong signal, such that e.g.~the wavefront can be determined. Moreover, it helps to obtain a minimal bias in \Xmax by avoiding having a dataset dominated by tens of thousands of antennas with essentially zero signal. 

Related to the detection threshold, and using the fact that the footprints generally form a contiguous area, we require that an antenna with a detection has a $\unit[25]{\%}$ detection rate in the nearby antennas in a radius of $\unit[20]{m}$. This avoids false positive detections outside the footprint area; tests have shown the results do not depend on the exact percentage or radius numbers.
A similar requirement was used at LOFAR, where half of the antennas in a 48-antenna station were required to have a detection for the station to be included \cite{Schellart:2014oaa}.

\subsection{Ensemble of simulated showers, fitting procedure}

We have produced an ensemble of 140 showers across 5 primary elements (50 H, 20 He, 20 C, 20 Si, 30 Fe), for each of 3 incoming directions. 
These are taken as arriving from East, for 3 zenith angles of 15, 30, and 40 degrees, respectively. 

Primary energy in the CoREAS simulations was set to $\unit[10^{17}]{eV}$, and the radio traces were scaled in amplitude proportional to the desired primary energy.
To evaluate the \Xmax reconstruction, we used each shower in turn as mock data, and reconstructed it using all other showers in the ensemble as model showers to fit to the data. The procedure was developed for LOFAR and is explained in more detail in \cite{Buitink:2014eqa}.

An example of a cosmic-ray radio footprint (primary energy of $\unit[10^{17}]{eV}$ at a zenith angle of $\unit[30]{^\circ}$ to be measured at SKA-Low is shown in Fig.~\ref{fig:footprint_example}.
Generally speaking, radio footprints are smaller for larger \Xmax values, i.e.~showers that peak closer to the ground.
However, there are many more subtle variations that a dense instrument such as SKA-Low is expected to pick up.

We fit simulated fluence footprints to data, where the core position is left free. To this end, we measure the noiseless fluence in the 208 simulated antennas on the radial grid, and interpolate the fluences thus found throughout the footprint using the method described in \cite{2023JInst..18P9005C}. This allows to produce fluences at the SKA antenna positions quickly when core positions are being varied in the fit procedure.

A chi-squared fit is done, fitting the fluence footprints of all 139 ensemble showers against the `data' shower. 
Free parameters in the fit are a scale factor $A$, used to measure the primary energy in real data, and the shower core position $(x_0, y_0)$, so the best fit is obtained by minimizing
\begin{equation}
\chi^2 = \sum_{\mathrm{antennas}} \left(\frac{A\,f_{\mathrm{model}}(x-x_0, y-y_0) - f_{\mathrm{data}}(x, y)}{\sigma_f(x, y)}\right)^2
\end{equation}
for each model shower, where $f_{\mathrm{model}}$ and $f_{\mathrm{data}}$ are the fluences per antenna in the model and data showers, respectively, and $\sigma_f$ is the uncertainty on the measured fluence due to noise.

The $\chi^2$-values will have a minimum as a function of the \Xmax of the model showers, close to the true \Xmax which we know for the simulated showers.
We fit a parabola to the lower envelope of points near the best-fitting shower, to obtain the \Xmax estimate.

\subsection{Using beamforming to increase signal-to-noise ratios}\label{sect:beamforming}
A limiting factor of the radio technique is a relatively high energy threshold, given by the requirement for the pulses to be detectable above the noise. 
The large number of antennas in the array, and in particular their high density, makes SKA well suited for analysis using interferometry and beamforming that improve the SNR and thus extend the measurable primary energy range downward. The motivation is to move from just below $\unit[10^{17}]{eV}$ down to around $\unit[10^{16}]{eV}$, and with newer techniques that would use full-array beamforming \cite{Scholten:2024upn,Schoorlemmer:2020low} even lower. This opens up an energy range previously unexplored by radio detection experiments.

For an air shower analysis as discussed here, the beamforming would be done offline, using the raw voltage traces of each antenna of which about $\unit[50]{\mu s}$ are stored. 
In contrast, in a distributed telescope like SKA or LOFAR, filtering and correlation of antenna pairs is typically done in real time for astronomical observations, as data volumes would be prohibitive otherwise. For air shower observations, both online beamforming and offline processing using the full nanosecond-level time resolution is feasible. 
Beamforming in any given direction is done by applying time shifts to each antenna's voltage trace, according to the expected geometric delays. Then, a sum over the antenna traces is a coherent addition and gives the \emph{beamformed trace}.

The cosmic-ray pulses arrive in a non-planar wavefront which in general has a hyperboloid shape~\cite{Corstanje:2014waa,Apel:2014usa}. This means that standard, far-field beamforming as usually applied in astronomy is not optimal.
Using near-field beamforming to points on the shower axis at various distances may become a practical reconstruction method in its own right \cite{Scholten:2024upn,Schoorlemmer:2020low,Schluter:2021egm} and will not be discussed here. 
A simpler option, in line with the method presented above, is to take patches of nearby antennas, e.g.~in groups of 4 or 16, and use far-field beamforming per group, as the baselines are short enough (on the order of \unit[3-10]{m}) for the wavefront curvature to be insignificant. The SNR would rise by a factor $\sqrt{n}$ when combining $n$ antennas.
This would give an effective number of measured signal traces a factor of 4 or 16 smaller, at 2 or 4 times the SNR, respectively.
We have emulated this process in simulations by taking one out of 4, 16, or 64 antennas, and boosting its SNR by the corresponding factor.

\section{Results of the reconstruction}
For the air shower reconstructions described here, we have used a decimation of 1 out of 4 antennas of every antenna station, as a conservative choice for the fraction of antennas that can be buffered.

\begin{figure}
	\centering
	\includegraphics[width=0.49\textwidth]{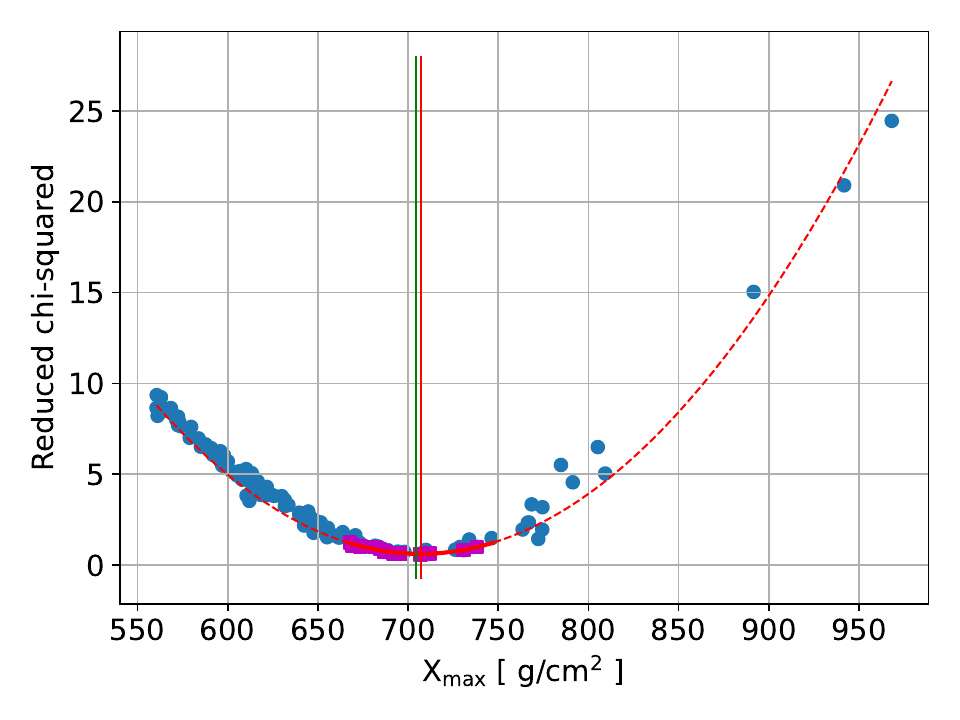}
        \includegraphics[width=0.49\textwidth]{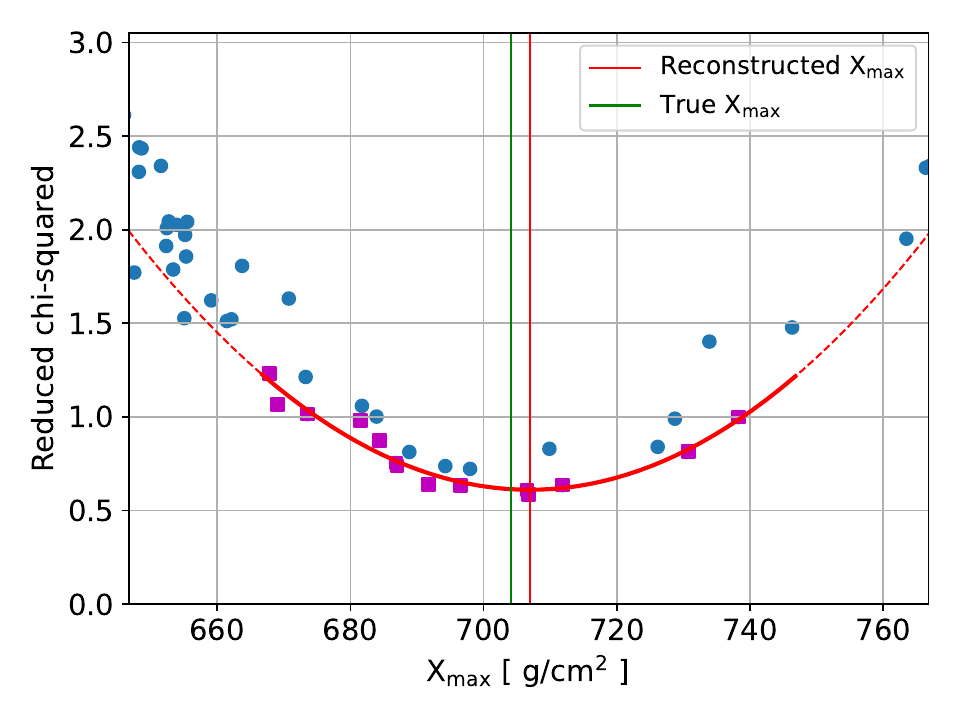}
        \caption{An example of an \Xmax reconstruction, showing the fit $\chi^2$ values versus \Xmax of each of the showers in the simulated ensemble, and a parabola fit to the lowest points around the minimum. The right panel shows a zoomed-in version of the same plot. Data points shown as magenta squares are the lower envelope of the data points and were used in the parabola fit.}
	\label{fig:xmax_reco_example}
\end{figure}

In Fig.~\ref{fig:xmax_reco_example}, we show an example of an \Xmax reconstruction of one shower in an ensemble of 140. The fit quality follows a curve with a minimum as a function of \Xmax which locally is well approximated by a parabola. To determine this parabola, we take an \Xmax range of $\pm \unit[40]{g/cm^2}$ around the \Xmax of the best-fitting shower. We select those fit quality points that form a lower envelope (magenta squares), and use them to fit the parabola. The condition for inclusion in the lower envelope is that for a given data point, lower chi-squared values are found in at most one direction, towards either higher or lower \Xmax. The motivation for using a lower envelope is that natural shower-to-shower variations introduce scatter in such a chi-squared plot even at a constant \Xmax, where lower values indicate better fits, with a lower limit for an `ideal' fit.

The \Xmax estimate is indicated, being the minimum of this parabola. The scatter is expected to stem from secondary parameters of the shower profile (not only \Xmax) and illustrate why novel reconstruction methods for SKA \cite{Corstanje:2023nlk,Corstanje:2023uyg} are believed to yield a more detailed insight. 

\subsection{Precision and bias of the \Xmax reconstruction}

By using each shower in turn as mock data, and reconstructing its \Xmax, we have evaluated the precision and bias on \Xmax, as a function of primary energy for this reconstruction procedure. We have removed the showers with the 5 highest and lowest \Xmax, i.e.~the lowest and highest $3.5$ percentile. In particular at high \Xmax, the density of showers is low which degrades the reconstruction precision. At LOFAR, we have treated this effectively by simulating extra (pre-selected) showers around a first estimate of \Xmax; in this analysis which uses a fixed ensemble, this is not feasible.

\begin{figure}
	\centering
	\includegraphics[width=0.49\textwidth]{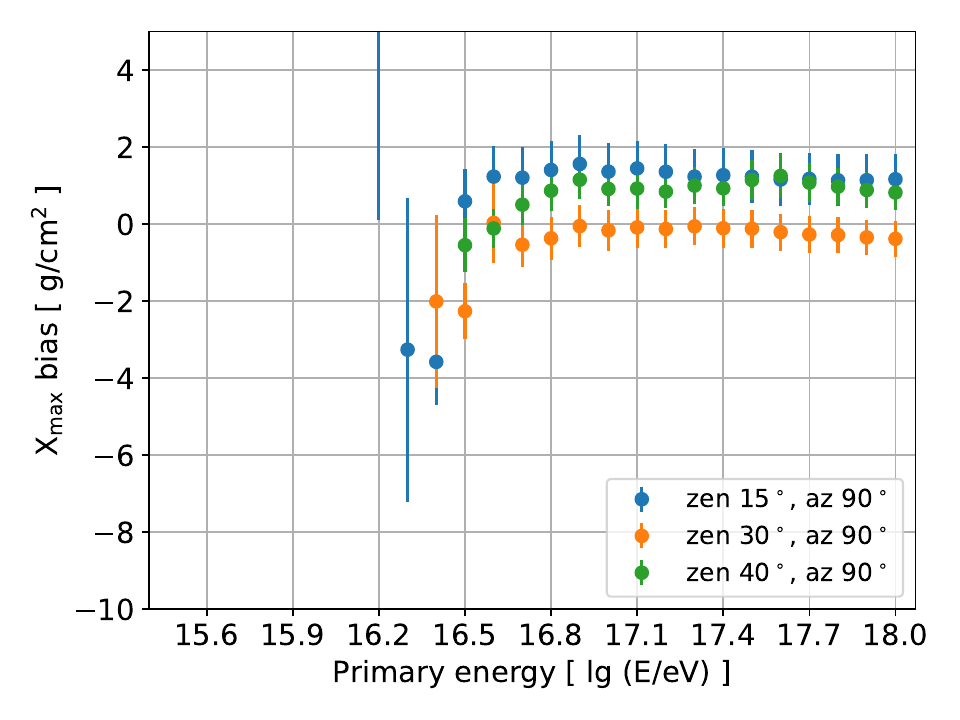}
        \includegraphics[width=0.49\textwidth]{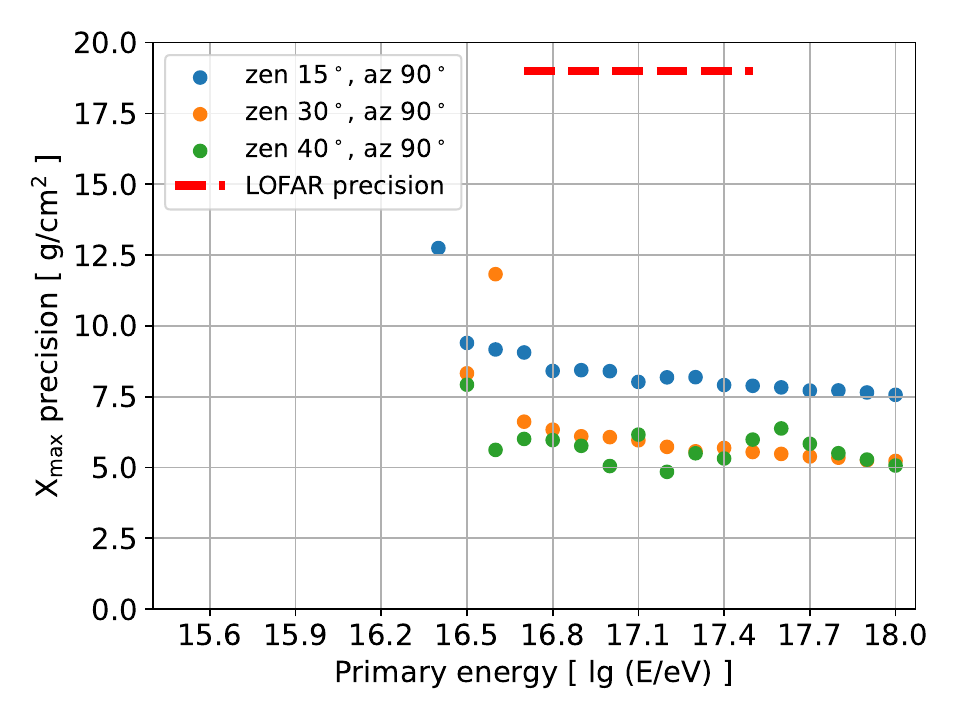}
	\caption{Bias (left panel) and precision (right panel) of the \Xmax reconstruction as a function of primary energy. Measurements were done with 1 out of 4 antennas of the SKA-Low array, using the SKALA4 antenna model. The primary energy range is the same as in Fig.~\ref{fig:xmax_bias_precision_beamforming} below which shows the additional energy range available when using beamforming. The precision attained at LOFAR is shown for comparison.}
	\label{fig:xmax_bias_precision}
\end{figure}

The results are shown in Fig.~\ref{fig:xmax_bias_precision}.
At primary energies of $\unit[10^{16.4}]{eV}$ and below, not all showers are reconstructed properly due to low SNR values. Hence, around this level, we find the low-energy cutoff for a reconstruction based on fluence measurements in single antennas.

At high primary energy, the precision, defined as the standard deviation of the \Xmax reconstruction errors, reaches a limiting value of about 5 to $\unit[8]{g/cm^2}$. Air showers at larger zenith angles are on average detectable in more antennas, due to the radio emission happening further away from the detector, hence `illuminating' more antennas. Therefore, the precision is somewhat worse for more vertical showers. 
For comparison, at LOFAR the median precision is $\unit[20]{g/cm^2}$, i.e.~the top of the range in Fig.~\ref{fig:xmax_bias_precision}, with only the highest-energy showers with a favorable core position reaching around $\unit[10]{g/cm^2}$.

We see that the precision is no longer limited by either the SNR value or the number of antennas. Instead, it reflects information in the air shower development that was neglected here and which is difficult to resolve using a limited simulation ensemble when focusing only on one parameter, the depth of its maximum \Xmax. For instance, the longitudinal evolution (along the shower track) can be modeled with two more parameters ($L$,$R$) with significant variations \cite{GaiserHillas,Matthews2010,Andringa:2011zz}. These parameters describe the width and asymmetry of the longitudinal distribution of particles, respectively. With an instrument like SKA-Low, these fine details will become accessible, opening up additional information on the mass composition that was not available earlier. Studies in this direction are ongoing \cite{Buitink:2023rso,Buitink:2023reh,Corstanje:2023nlk,Corstanje:2023uyg}.

The bias on \Xmax, i.e.~the average reconstruction mismatch over the 140 showers, is less than $\unit[1.5]{g/cm^2}$ which is well below other systematic uncertainties such as those arising from the hadronic interaction models and errors in the local atmosphere models at the time of the showers \cite{Mitra:2020mza}. In principle, potential uncertainties from the simulation process itself would add to the systematic error budget, but as e.g.~the radio emission is calculated deterministically from Maxwell's equations, giving consistent results across simulation codes (also discussed in \cite{Corstanje:2021kik}), the above sources of uncertainty are expected to dominate.

Nevertheless, it is important to be aware of any source of bias on \Xmax and lower it where possible. After all, reducing the systematic error budget as much as possible is key to improving the mass composition results, because as shown in Sect.~\ref{sect:masscomposition}, at a reasonably attainable number of measurements, the results become limited by systematics, not statistics.

We have also run the same analysis with the older SKALA2 antenna model, yielding no significant differences in the performance in terms of bias, precision and low-energy limit.
It follows that the performance of the method is not sensitive to the finer details of the antenna response, as long as it is simulated or measured accurately.
We conclude that the given reconstruction method used at LOFAR still performs well when reconstructing \Xmax only, in a setting with tens of thousands of antennas at SKA-Low, even when not yet optimized for the wider frequency bandwidth.

\subsection{Low-energy showers, beamforming}
\begin{figure}
	\centering
	\includegraphics[width=0.49\textwidth]{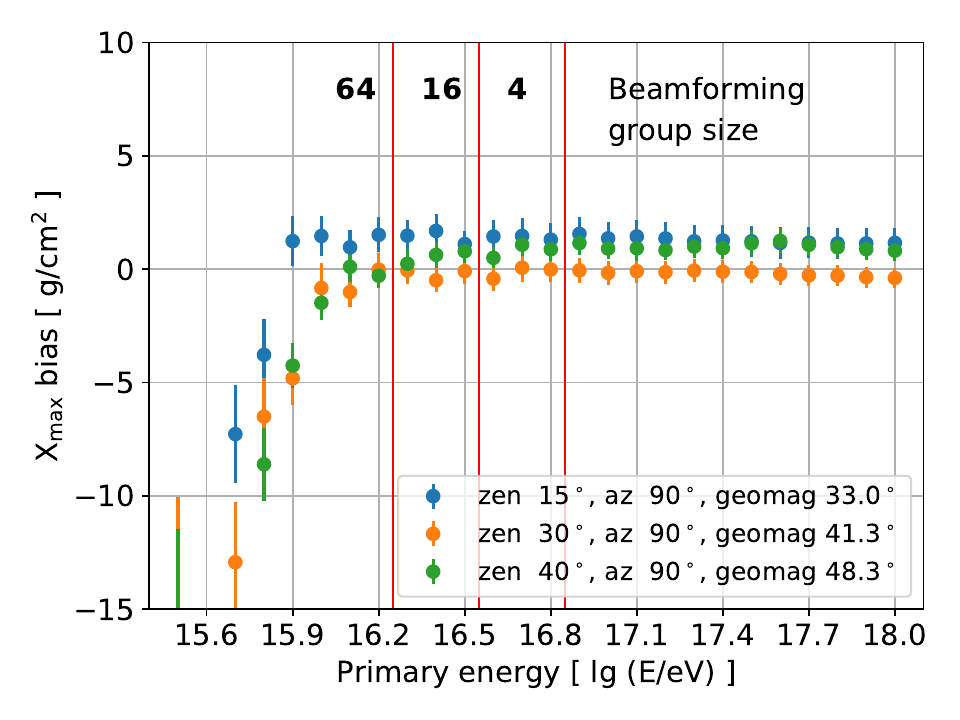}
        \includegraphics[width=0.49\textwidth]{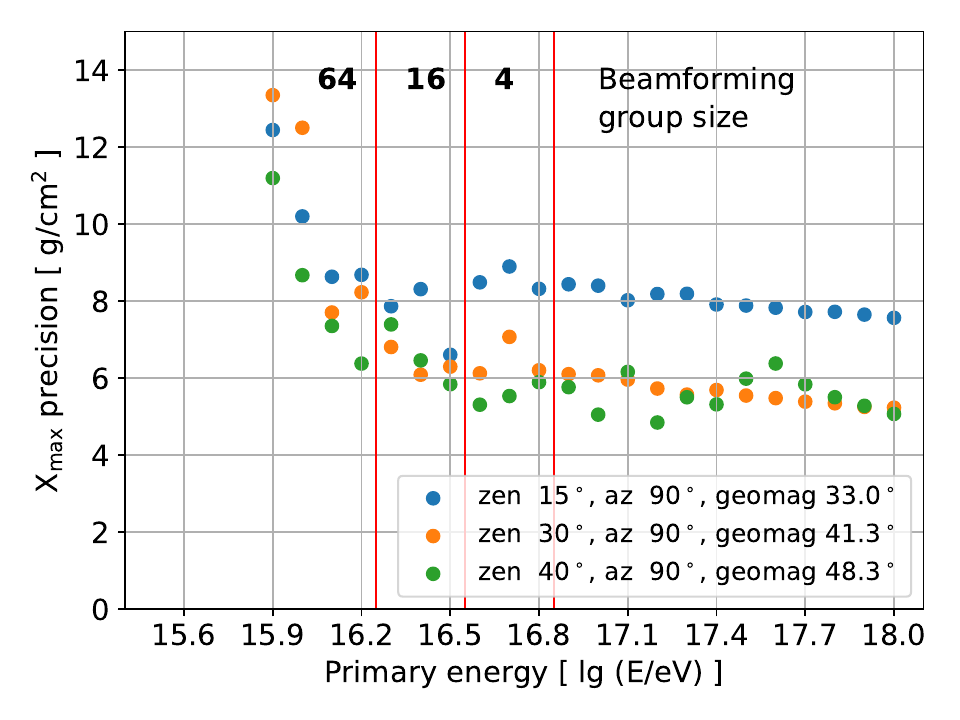}
	\caption{Bias (left panel) and precision (right panel) of \Xmax reconstruction as a function of primary energy. Red vertical lines indicate thresholds below which beamforming is applied, reducing the number of effective antennas by the given factor while increasing signal-to-noise ratios. Again, 1 in 4 antennas of the full array were taken as a baseline. }
	\label{fig:xmax_bias_precision_beamforming}
\end{figure}

To extend the accessible energy range downward, we mimicked the process of patch-wise beamforming as outlined in Sect.~\ref{sect:beamforming} by only using 1 in $N$ antennas, where $N=4,16,64$ at an SNR increased by a factor $2,4,8$, respectively. Again, the baseline array was taken as 1 out of 4 antennas of the SKA-Low core.
The precision and bias results for \Xmax are shown in Fig.~\ref{fig:xmax_bias_precision_beamforming}, with a stepwise increase in beamforming level towards lower energy. The thresholds are indicated by the vertical lines, chosen somewhat arbitrarily to obtain useful results for this study.
The results indicate that \Xmax reconstructions can be reliably done for energies down to $\unit[10^{16}]{eV}$. Again, more sophisticated methods based on beamforming are being developed and will likely improve on these results, extending the available energy range down to well below $\unit[10^{16}]{eV}$.

\subsection{Restrictions in dynamic range}

\begin{figure}
	\centering
	\includegraphics[width=0.49\textwidth]{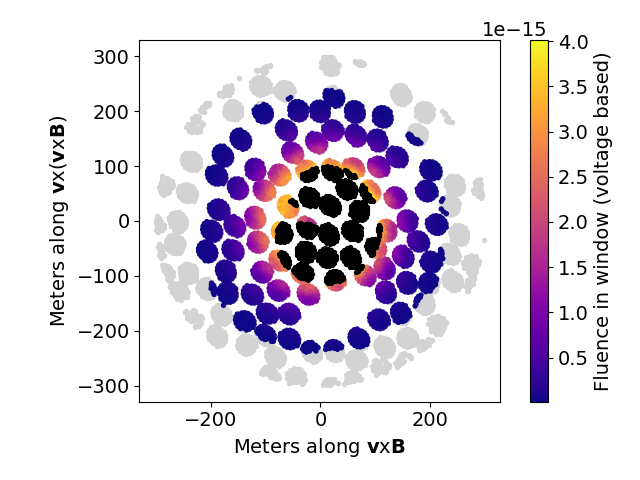}
       \includegraphics[width=0.49\textwidth]{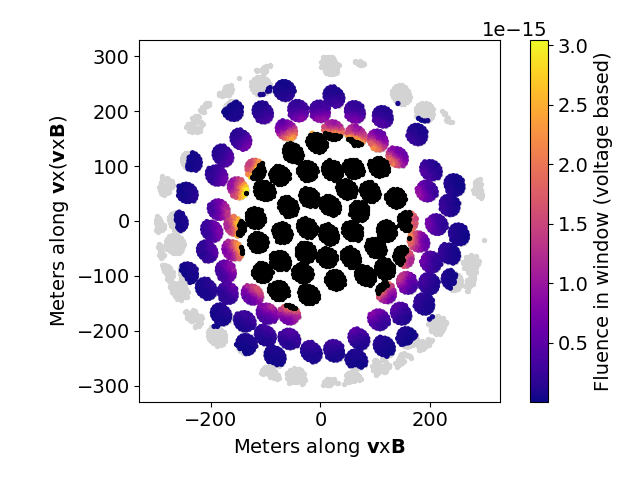}
       
	\caption{Two examples of a measured radio footprint of a cosmic-ray air shower, with a dynamic range limited to $\unit[32]{\sigma}$.
    Left panel: primary energy $\unit[10^{17.5}]{eV}$; right panel: $\unit[10^{17.8}]{eV}$. The antennas excluded due to exceeding the dynamic range (in at least one polarization) are marked as black dots and are found in the central region of the footprint. }
	\label{fig:footprint_example_dynrange}
\end{figure}

\begin{figure}
	\centering
	\includegraphics[width=0.49\textwidth]{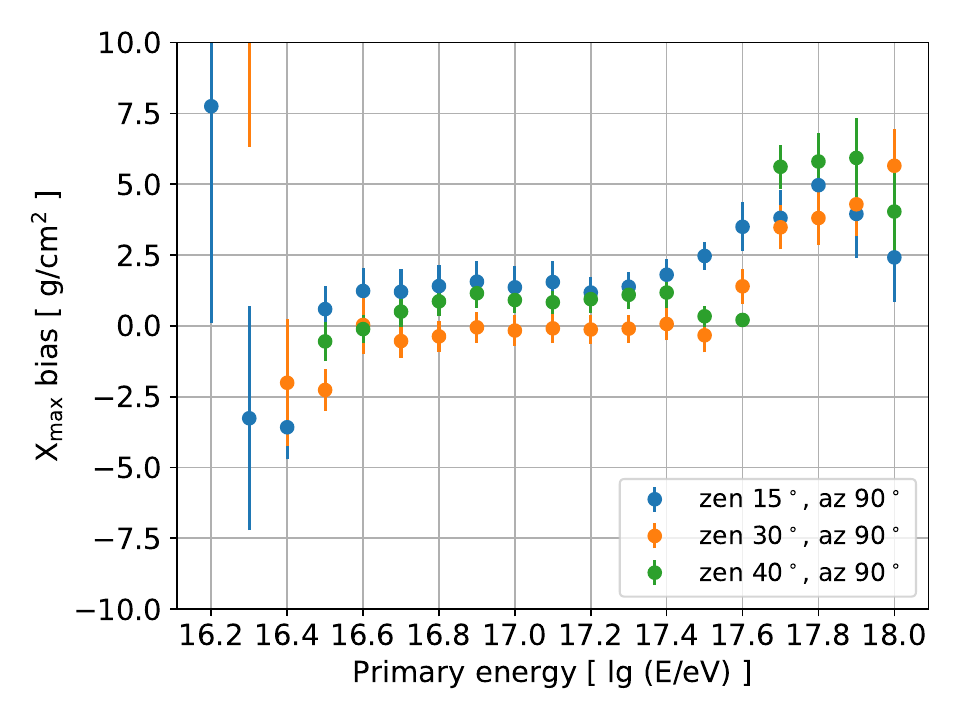}
        \includegraphics[width=0.49\textwidth]{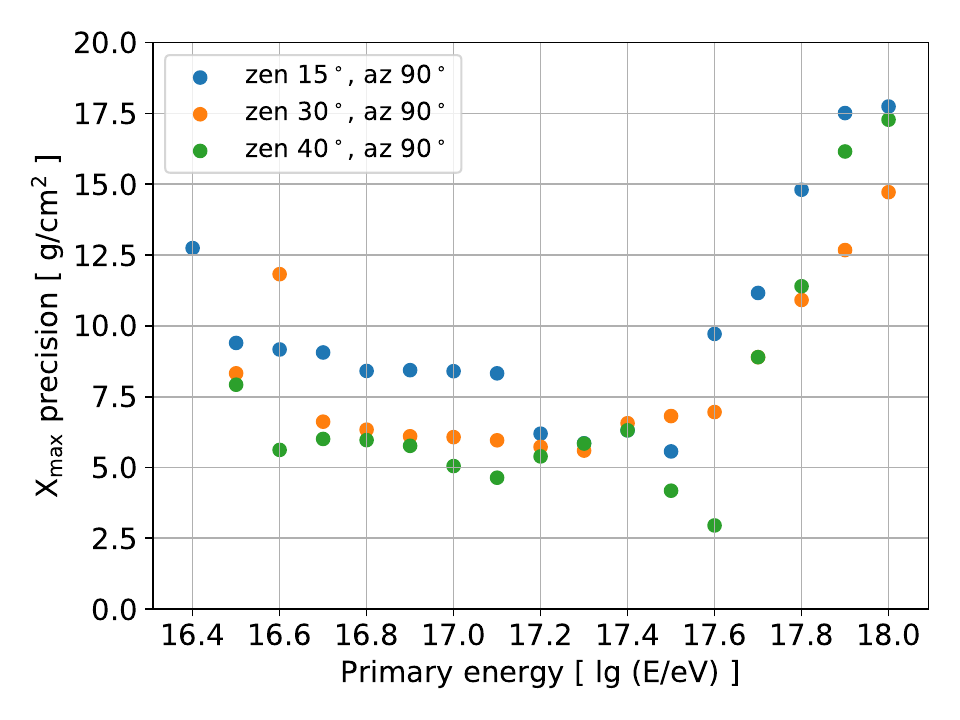}
	\caption{Bias (left panel) and precision (right panel) of \Xmax reconstruction as a function of primary energy. Here, a dynamic range restriction to $\unit[32]{\sigma}$ limits the accuracy above $\mathrm{lg}(E/\unit{eV}) \approx 17.4$ as compared to Fig.~\ref{fig:xmax_bias_precision}. }
	\label{fig:xmax_bias_precision_dynamicrange}
\end{figure}

The dynamic range of the SKA-Low instrument is currently being refined. As of now, it is anticipated that the digitization of the voltage signals is done using 8-bit analog-to-digital converters (ADC) \cite{SKA-documents}. The signals from all directions on the sky, which in a single wide-view antenna are all detected simultaneously as `noise', will have their one-sigma level inside this range. We assume this to be at the second-to-least significant bit or higher.
Only then, correlating and/or beamforming the signals can be performed to successfully image the astronomical sources.
For comparatively large signals, like those from air showers, this however sets a hard limit on the maximal pulse strength that can reliably be detected without losing information. 

The level at which the noise appears in the ADCs is a setting under control of the observatory, by using adjustable attenuators in the signal chain.
Both the absolute level and the control settings will vary during the day. One of the considerations is impulsive radio frequency interference (RFI) that may cause spurious broadband signals when it is clipped (exceeds maximum ADC level). Thus, the final design is still being discussed. 

To see how these settings would impact cosmic-ray measurements, in particular in the higher primary energy range, we will as an example assume the one-sigma level is set at the third-lowest bit, i.e.~ADC value 4. This means signals with a peak level stronger than $\unit[32]{\sigma}$ saturate the ADCs and will be clipped. 
While there exist mitigation strategies such as de-clipping algorithms, here we remove each antenna in which at least one polarization has a clipped signal.
This is a rather restrictive criterion; in a more refined analysis one could use the weakest polarization if it is not clipped, and a closer look at the leading edge of the pulse shape could allow a measurement before clipping sets in.
For the present analysis, it is sufficient to obtain a reasonable lower bound on what may be achieved. 

An example footprint for a proton shower at a zenith angle of $\unit[30]{^\circ}$ and at primary energies $10^{17.5}$ and $\unit[10^{17.8}]{eV}$ is shown in Fig.~\ref{fig:footprint_example_dynrange}. Antennas with a clipped signal are shown as black dots. Effectively, only the outer, weaker part of the footprint is being measured here, although due to the higher energy, many antennas are still present that contain a useful signal.

The results for the \Xmax measurements when we exclude all antennas with saturated signals are shown in Fig.~\ref{fig:xmax_bias_precision_dynamicrange}, which may be compared to those in Fig.~\ref{fig:xmax_bias_precision}. Above a primary energy about $\unit[10^{17.5}]{eV}$, both bias and precision tend to worsen.
Nevertheless, the precision remains better than $\unit[20]{g/cm^2}$ which is the typical precision obtained at LOFAR (although there on average for lower energies). The bias increasing to a level near $\unit[5]{g/cm^2}$ is somewhat more concerning, and should be investigated further together with other mitigation strategies for ADC-saturating signals.

The observatory settings for the noise level determine the energy level beyond which the results degrade; each ADC-bit lower or higher corresponds to $0.3$ in log-energy higher or lower, respectively.
The degrading effects on the \Xmax measurements appear manageable, even with the rather harsh removal of all clipped antenna traces. However, it is likely that measuring finer details of the shower evolution will be impacted more strongly, in particular when important shower information is present in the inner region around the shower core. This remains an important point for further study, especially when realistic observatory settings become clear in practice. 
\section{Towards cosmic-ray mass composition measurements at SKA-Low}\label{sect:masscomposition}
For a measurement of the mass composition in a given energy range, many \Xmax measurements are combined in a statistical analysis, effectively comparing the measured \Xmax distribution with simulated distributions of \Xmax combining different fractions of primary nuclei. 

There are two common approaches, of which the feasibility depends on the number of measured showers.
The easiest is to consider the average \Xmax as a function of energy, which to first order corresponds to heavier (lower \Xmax) or lighter (higher \Xmax) mass composition.
Similarly, the second moment of the \Xmax distribution can be estimated, from the standard deviation of \Xmax. This requires a somewhat larger number of measurements.

The most precise estimate of the mass composition is obtained by comparing a mixed-composition model of several primary elements to the data, at distribution level.
This exploits all the information in the measured \Xmax and energy values.
It requires many measured showers while keeping systematic uncertainties as low as possible; improvements on previous LOFAR results using 334 showers \cite{Corstanje:2021kik} are possible and desirable.

\subsection{Impact of statistical and systematic uncertainties}

For LOFAR, systematic uncertainties of about 7 to $\unit[9]{g/cm^2}$ on the reconstructed average \Xmax were used when comparing to the results by other experiments. The contributions that combine into these numbers are given in \cite{Corstanje:2021kik}.
The number of measurements to aim for SKA-Low should therefore be guided by having statistical uncertainties much smaller than the systematics on such estimates.
As the standard deviation of e.g.~the \Xmax distribution for protons is roughly $\unit[60]{g/cm^2}$, having $N=1000$ showers per energy bin would give statistical uncertainties of $\unit[2]{g/cm^2}$, satisfying this requirement.
From a bootstrap analysis, uncertainties on the estimate of the standard deviation of the \Xmax distribution would then be about $\unit[3]{g/cm^2}$, which is also satisfactory. 

To evaluate the statistical and systematic uncertainties on the mass composition, we have done a parametric bootstrap analysis, creating \Xmax datasets by sampling from a mixed-composition \Xmax distribution.
To this end, we have taken the best-fit mass composition in \cite{Corstanje:2021kik}, quoted as $\unit[28]{\%}$ hydrogen, $\unit[11]{\%}$ helium, $\unit[60]{\%}$ nitrogen (as a proxy for C/N/O) and $\unit[1]{\%}$ iron, and assumed the Sibyll-2.3d hadronic interaction model \cite{Riehn:2019jet}.
In a parametric bootstrap run, the best-fit values as well as the uncertainties naturally vary from one drawing to another. We have chosen a run that produces a best fit reasonably close to the quoted values, in order to compare the uncertainty bars across datasets. This allows to assess at least qualitatively at which sample size the estimates are systematics-limited. 

\begin{figure}[h]
	\centering
	\includegraphics[width=0.49\textwidth]{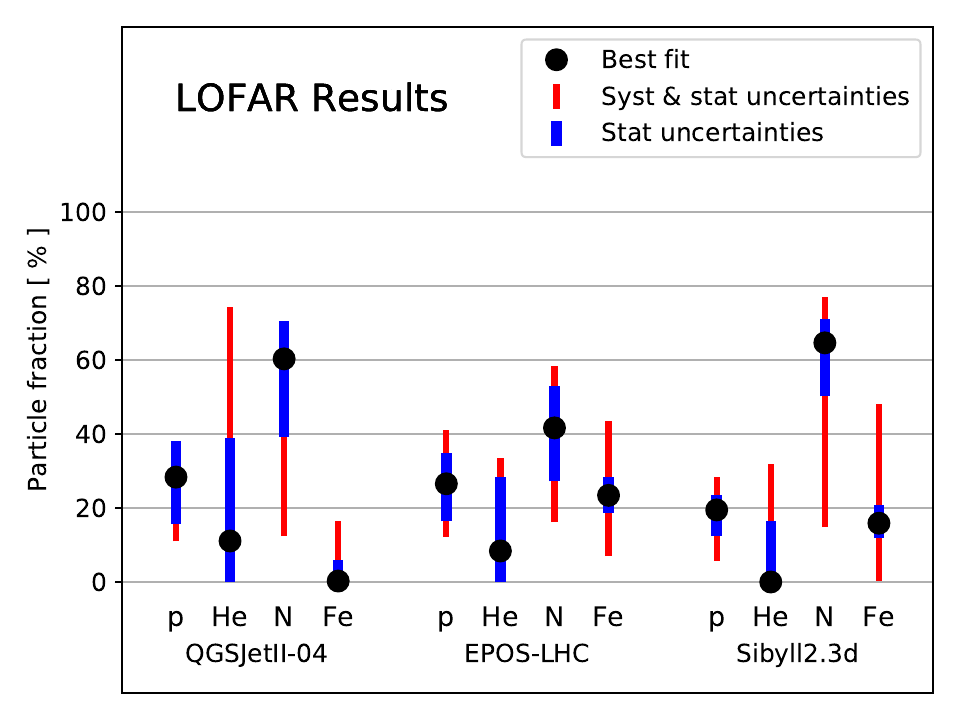}
    \includegraphics[width=0.49\textwidth]{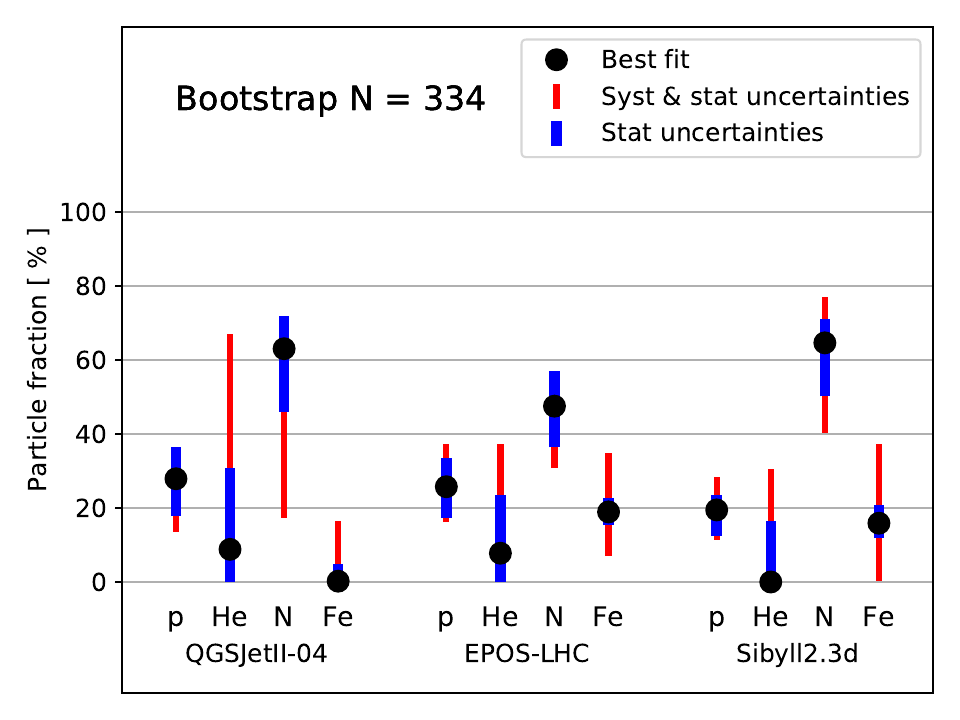}
       \includegraphics[width=0.49\textwidth]{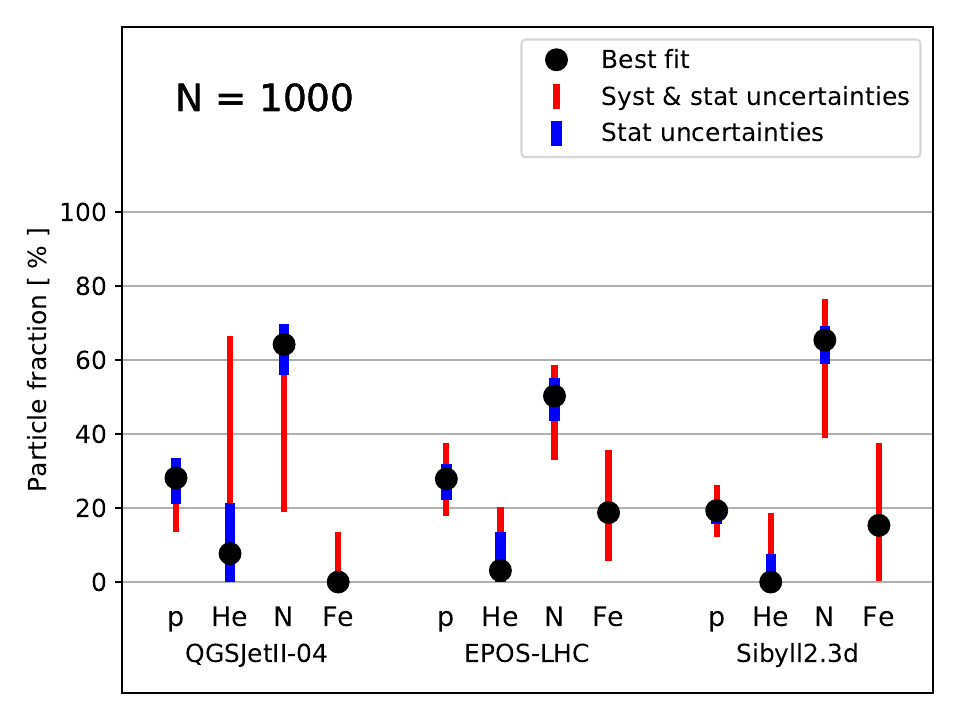}
       \includegraphics[width=0.49\textwidth]{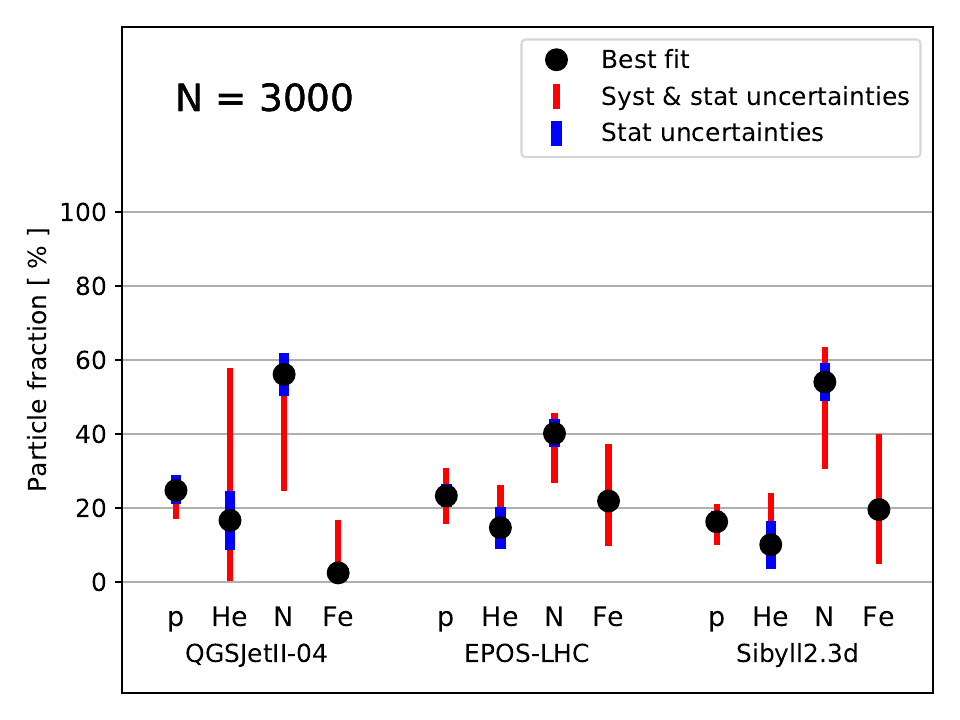}
       
	\caption{The LOFAR mass composition result based on 334 showers, and example mass composition estimates from a bootstrapped sample, of $N=334$, $1000$, and $3000$ detected with SKA-Low, respectively. This gives an impression of how the statistical and systematic uncertainties evolve with increasing number of showers.
    For large datasets, a mass composition analysis from \Xmax only is inherently systematics-limited; using additional information in the air showers this is expected to be improved further.}
\label{fig:bootstrap_composition_results}
\end{figure}

We have evaluated the mass composition and its uncertainties for a simulated $N=334$, $N=1000$ and an $N=3000$ dataset of SKA, accounting for a total systematic uncertainty on \Xmax of $\unit[9]{g/cm^2}$ as in the LOFAR case. Statistical uncertainties on \Xmax were set to $\unit[7]{g/cm^2}$ in line with the results found above, and the energy to a constant of $\unit[10^{17}]{eV}$ anticipating a narrow energy binning. The distribution shapes do not change much at lower or higher energies, apart from their average and minor changes to the standard deviation.
The results are shown in Fig.~\ref{fig:bootstrap_composition_results} where, as in the LOFAR analysis, the statistical analysis has been done assuming the \Xmax probability distributions from three hadronic interaction models. Best-fit values and statistical errors are shown; the extended (red) uncertainties are obtained from shifting all \Xmax values by their systematic uncertainty i.e.~$\pm \unit[9]{g/cm^2}$, and redoing the analysis which again includes statistical errors. 

The uncertainties on the run with 334 showers are noticeably smaller than those on the LOFAR data, but not by much. The difference arises from the lower statistical errors on \Xmax, i.e.~7 instead of on average $\unit[20]{g/cm^2}$.
For 3000 or more showers, the results are purely limited by systematics, with statistical errors becoming negligible. Given this toy study, we conclude that 1000 to 3000 showers per energy bin are sufficient for a purely \Xmax-based mass composition analysis.
However, this also points out the need to include more shower information in the reconstructions, beyond \Xmax, as for instance distinguishing hydrogen and helium is quite relevant for astrophysical models of cosmic-ray production, but always challenging using \Xmax alone. As discussed above, we anticipate SKA to be one of the first experiments to be sensitive to more subtle structures of the shower profile. Methods to this end are currently under development \cite{Buitink:2024/j}. 

\subsection{Natural and technical constraints to the number of observed cosmic-ray air showers}
Cosmic rays exhibit a steeply falling energy spectrum \cite{ParticleDataGroup:2024cfk}; its spectral index near $-3$ indicates that the number of cosmic rays in a primary energy binned in log-energy around $\unit[10^{18}]{eV}$ is roughly 100 times lower than at $\unit[10^{17}]{eV}$. Therefore, our observable range is limited from above by the scarcity of cosmic rays at very high energies, while at the lower end of the range accessible to a radio telescope, technical limitations are likely to arise in the number of showers that can be recorded and processed.

Measured showers are usually binned in log-energy, analyzing \Xmax statistics and mass composition in each bin to see trends with energy.
Whenever the number of measured showers allows, a bin width of $0.1$ in $\mathrm{lg}\, (E/\mathrm{eV})$ is used (see e.g.~\cite{PierreAuger:2024flk}) to obtain a detailed spectrum.
The primary energy resolution will be better than $\unit[9]{\%}$ as was found from LOFAR analysis \cite{Corstanje:2021kik}, or $0.04$ in log-energy. With systematic errors of $\unit[14]{\%}$ or better (again from LOFAR analysis), translating to $0.057$ in log-energy, we see that a log-energy bin width of $0.1$ is well attainable and also should not be lowered further.

We can integrate the power-law spectrum from \cite{ParticleDataGroup:2024cfk} for an energy bin of for example $\mathrm{lg}\, (E/\mathrm{eV}) = 17.0$ to $17.1$, to obtain the expected number of showers in an observing year, i.e., one year of effective up-time of the cosmic-ray observing mode as it runs in parallel with the astronomical observations.

\begin{figure}
        \centering
        \includegraphics[width=0.49\textwidth]{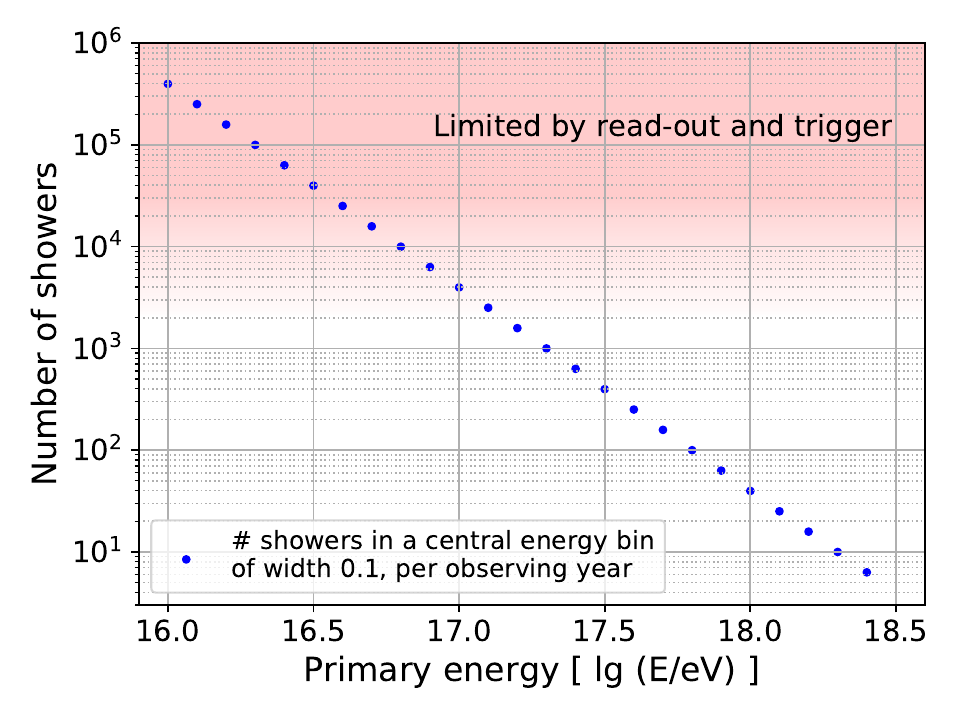}  
	\caption{Estimate of expected event counts per SKA-Low observing year, in an energy bin of width $0.1$ in $\mathrm{lg} (E/\mathrm{eV})$ centered on the given primary energy. This approximates the spectrum by a constant $E^{-3}$ power law, and does not include any quality cuts. The shaded region indicates a technical limit to the shower triggering and data transfer rates that will arise from running in parallel alongside ongoing astronomical observations. It is again expressed in terms of number of showers per bin, as it can be expected at order-of-magnitude level.}
	\label{fig:cr_energy_spectrum}
\end{figure}

Using the area of the inner part of SKA-Low of about $\unit[0.8]{km^2}$ and a zenith angle range of 0 to $\unit[55]{^\circ}$ which is $\unit[2.68]{sr}$, we obtain about $3\times 10^3$ showers in the energy bin, or up to $8\times 10^3$ showers in total above $\unit[10^{17}]{eV}$.
In Fig.~\ref{fig:cr_energy_spectrum}, we show the expected number of showers per observing year, in a log-energy bin of width 0.1, as a function of primary energy.
We conclude that for one observing year, the sample will be limited to less than 1000 showers per energy bin above about $\mathrm{lg}\, (E/\mathrm{eV}) = 17.3$. This illustrates the need for long observing times and/or a large collecting area, which all cosmic-ray experiments share. 
While, as shown above, roughly 1000 showers per energy bin is optimal, lower numbers give already useful estimates. At the higher end of the spectrum, energy bins are usually widened to increase the level of statistics at the expense of some energy resolution.

At low energies, the particle detectors may no longer reach full efficiency in triggering every shower.
The threshold energy where full efficiency is reached depends on details that are still unknown, such as the triggering logic and the spacing of the detectors, as well as their size and readout. Experience from the comparable triggering array at LOFAR, yields a full efficiency threshold for all primaries and all arrival directions at \unit{$10^{16.6}$}{eV} \cite{Thoudam:2015lba} and roughly $\unit[50]{\%}$ a decade lower in energy. 

Using $\unit[10^{17.3}]{eV}$ as benchmark energy and obtaining 1000 showers per bin below this energy and roughly 2000 showers above, would lead to roughly 15,000 showers in an observing year. This naturally does not include any quality or anti-bias cuts, but provides an estimate to study the burden on the system. For this, we consider the requirements for triggering and network bandwidth at order-of-magnitude level.
The uncompressed data volume $S$ works out to
\begin{equation}
    S \approx \unit[75]{TB}\, \left(\frac{t_{\mathrm{obs}}}{\unit[1]{yr}}\right)\,\left(\frac{\Delta t_{\mathrm{trace}}}{\unit[50]{\mu s}}\right)\,\left(\frac{N_{\mathrm{ant}}}{60000}\right),
\end{equation}
when reading traces of length $\Delta t_{\mathrm{trace}}$ at $\unit[800]{MHz}$ and 8 bits per sample.
Per measured air shower, this amounts to $\unit[4.5]{GB}$, which has to be transferred in parallel to ongoing observations. At LOFAR $\unit[2.1]{ms}$ are read out, which were chosen for a high-precision characterization of the RFI environment. $\unit[50]{\mu s}$ is, however, already longer than the typical air shower experiment which records $\unit[10]{\mu s}$ or less, with a more complex triggering logic that centers the pulse. Therefore, the data volume could be decreased at the expense of a more complex trigger and a more limited background noise characterization.

For 15,000 showers per year, we would have to trigger roughly twice per hour on average, keeping in mind that the time between events is random and follows an exponential distribution.
This then means that the buffer readout has to happen in a small fraction of this, i.e.~on the order of a minute or less. Reading out the buffer is dead time for any transient buffer observing mode, unless for instance a double buffering system is used.
A corresponding data rate of at least $\unit[80]{MB/s}$ from many station-level buffers in parallel appears reasonable, although the practical throughput depends on hardware details.

\section{Summary}
To estimate the potential of detecting air showers with the Square Kilometre Array, we have applied the current state-of-the-art reconstruction methods as developed for LOFAR to dedicated air shower simulations for SKA. In particular, we have investigated the obtainable reconstruction performance of the shower maximum, which has been found to be as precise as 5 to $\unit[8]{g/cm^2}$ between $\unit[10^{16.6}]{eV}$ and $\unit[10^{18}]{eV}$, with a negligible bias. We furthermore investigated how taking full advantage of the 57,000 antennas of SKA-Low by beamforming can lower the energy threshold to $\unit[10^{16}]{eV}$ without introducing significant biases or losing significant resolution. A potentially limited dynamic range will, however, impact the obtainable resolution at energies exceeding about $\unit[10^{17.5}]{eV}$, which should be a consideration for the SKA systems design, in addition to the impact of natural and man-made impulsive RFI on a system with a limited dynamic range. Lastly, we found that SKA is likely to reach a limit in mass composition studies when using only the depth of shower maximum \Xmax. Work is therefore required and ongoing to develop methods that reconstruct the shower profile in more detail to obtain a more direct handle on the cosmic-ray composition. An instrument like SKA-Low is well equipped to advance the field in this direction. 

\section*{Acknowledgments}
SBo, AN, KT acknowledge the Verbundforschung of the German Ministry for Education and Research (BMBF). 
PL and KW are supported by the Deutsche Forschungsgemeinschaft (DFG, German Research Foundation) – Projektnummer 531213488.
MD is supported by the Flemish Foundation for Scientific Research (FWO-AL991).
ST acknowledges funding from the Khalifa University RIG-S-2023-070 grant.
The authors gratefully acknowledge the computing time provided on the high-performance computer HoreKa by the National High-Performance Computing Center at KIT (NHR@KIT). This center is jointly supported by the Federal Ministry of Education and Research and the Ministry of Science, Research and the Arts of Baden-Württemberg, as part of the National High-Performance Computing (NHR) joint funding program. HoreKa is partly funded by the German Research Foundation.

\bibliography{ska_xmax}

\begin{thebibliography}{55}%
\makeatletter
\providecommand \@ifxundefined [1]{%
 \@ifx{#1\undefined}
}%
\providecommand \@ifnum [1]{%
 \ifnum #1\expandafter \@firstoftwo
 \else \expandafter \@secondoftwo
 \fi
}%
\providecommand \@ifx [1]{%
 \ifx #1\expandafter \@firstoftwo
 \else \expandafter \@secondoftwo
 \fi
}%
\providecommand \natexlab [1]{#1}%
\providecommand \enquote  [1]{``#1''}%
\providecommand \bibnamefont  [1]{#1}%
\providecommand \bibfnamefont [1]{#1}%
\providecommand \citenamefont [1]{#1}%
\providecommand \href@noop [0]{\@secondoftwo}%
\providecommand \href [0]{\begingroup \@sanitize@url \@href}%
\providecommand \@href[1]{\@@startlink{#1}\@@href}%
\providecommand \@@href[1]{\endgroup#1\@@endlink}%
\providecommand \@sanitize@url [0]{\catcode `\\12\catcode `\$12\catcode
  `\&12\catcode `\#12\catcode `\^12\catcode `\_12\catcode `\%12\relax}%
\providecommand \@@startlink[1]{}%
\providecommand \@@endlink[0]{}%
\providecommand \url  [0]{\begingroup\@sanitize@url \@url }%
\providecommand \@url [1]{\endgroup\@href {#1}{\urlprefix }}%
\providecommand \urlprefix  [0]{URL }%
\providecommand \Eprint [0]{\href }%
\providecommand \doibase [0]{https://doi.org/}%
\providecommand \selectlanguage [0]{\@gobble}%
\providecommand \bibinfo  [0]{\@secondoftwo}%
\providecommand \bibfield  [0]{\@secondoftwo}%
\providecommand \translation [1]{[#1]}%
\providecommand \BibitemOpen [0]{}%
\providecommand \bibitemStop [0]{}%
\providecommand \bibitemNoStop [0]{.\EOS\space}%
\providecommand \EOS [0]{\spacefactor3000\relax}%
\providecommand \BibitemShut  [1]{\csname bibitem#1\endcsname}%
\let\auto@bib@innerbib\@empty
\bibitem [{\citenamefont {Aab}\ \emph {et~al.}(2016{\natexlab{a}})\citenamefont
  {Aab} \emph {et~al.}}]{PierreAuger:2015hbf}%
  \BibitemOpen
  \bibfield  {author} {\bibinfo {author} {\bibfnamefont {A.}~\bibnamefont
  {Aab}} \emph {et~al.} (\bibinfo {collaboration} {Pierre Auger}),\ }\bibfield
  {title} {\bibinfo {title} {{Energy Estimation of Cosmic Rays with the
  Engineering Radio Array of the Pierre Auger Observatory}},\ }\href
  {https://doi.org/10.1103/PhysRevD.93.122005} {\bibfield  {journal} {\bibinfo
  {journal} {Phys. Rev. D}\ }\textbf {\bibinfo {volume} {93}},\ \bibinfo
  {pages} {122005} (\bibinfo {year} {2016}{\natexlab{a}})},\ \Eprint
  {https://arxiv.org/abs/1508.04267} {arXiv:1508.04267 [astro-ph.HE]}
  \BibitemShut {NoStop}%
\bibitem [{\citenamefont {Aab}\ \emph {et~al.}(2016{\natexlab{b}})\citenamefont
  {Aab} \emph {et~al.}}]{PierreAuger:2016vya}%
  \BibitemOpen
  \bibfield  {author} {\bibinfo {author} {\bibfnamefont {A.}~\bibnamefont
  {Aab}} \emph {et~al.} (\bibinfo {collaboration} {Pierre Auger}),\ }\bibfield
  {title} {\bibinfo {title} {{Measurement of the Radiation Energy in the Radio
  Signal of Extensive Air Showers as a Universal Estimator of Cosmic-Ray
  Energy}},\ }\href {https://doi.org/10.1103/PhysRevLett.116.241101} {\bibfield
   {journal} {\bibinfo  {journal} {Phys. Rev. Lett.}\ }\textbf {\bibinfo
  {volume} {116}},\ \bibinfo {pages} {241101} (\bibinfo {year}
  {2016}{\natexlab{b}})},\ \Eprint {https://arxiv.org/abs/1605.02564}
  {arXiv:1605.02564 [astro-ph.HE]} \BibitemShut {NoStop}%
\bibitem [{\citenamefont {Abdul~Halim}\ \emph
  {et~al.}(2024{\natexlab{a}})\citenamefont {Abdul~Halim} \emph
  {et~al.}}]{Auger_energy:2024}%
  \BibitemOpen
  \bibfield  {author} {\bibinfo {author} {\bibfnamefont {A.}~\bibnamefont
  {Abdul~Halim}} \emph {et~al.} (\bibinfo {collaboration} {Pierre Auger
  Collaboration}),\ }\bibfield  {title} {\bibinfo {title} {{Demonstrating
  Agreement between Radio and Fluorescence Measurements of the Depth of Maximum
  of Extensive Air Showers at the Pierre Auger Observatory}},\ }\href
  {https://doi.org/10.1103/PhysRevLett.132.021001} {\bibfield  {journal}
  {\bibinfo  {journal} {Phys. Rev. Lett.}\ }\textbf {\bibinfo {volume} {132}},\
  \bibinfo {pages} {021001} (\bibinfo {year} {2024}{\natexlab{a}})}\BibitemShut
  {NoStop}%
\bibitem [{\citenamefont {Buitink}\ \emph {et~al.}(2014)\citenamefont {Buitink}
  \emph {et~al.}}]{Buitink:2014eqa}%
  \BibitemOpen
  \bibfield  {author} {\bibinfo {author} {\bibfnamefont {S.}~\bibnamefont
  {Buitink}} \emph {et~al.} (\bibinfo {collaboration} {LOFAR}),\ }\bibfield
  {title} {\bibinfo {title} {{Method for high precision reconstruction of air
  shower $X_{max}$ using two-dimensional radio intensity profiles}},\ }\href
  {https://doi.org/10.1103/PhysRevD.90.082003} {\bibfield  {journal} {\bibinfo
  {journal} {Phys. Rev. D}\ }\textbf {\bibinfo {volume} {90}},\ \bibinfo
  {pages} {082003} (\bibinfo {year} {2014})},\ \Eprint
  {https://arxiv.org/abs/1408.7001} {arXiv:1408.7001 [astro-ph.IM]}
  \BibitemShut {NoStop}%
\bibitem [{\citenamefont {Abdul~Halim}\ \emph
  {et~al.}(2024{\natexlab{b}})\citenamefont {Abdul~Halim} \emph
  {et~al.}}]{Auger_radio:2024}%
  \BibitemOpen
  \bibfield  {author} {\bibinfo {author} {\bibfnamefont {A.}~\bibnamefont
  {Abdul~Halim}} \emph {et~al.} (\bibinfo {collaboration} {Pierre Auger
  Collaboration}),\ }\bibfield  {title} {\bibinfo {title} {{Radio measurements
  of the depth of air-shower maximum at the Pierre Auger Observatory}},\ }\href
  {https://doi.org/10.1103/PhysRevD.109.022002} {\bibfield  {journal} {\bibinfo
   {journal} {Phys. Rev. D}\ }\textbf {\bibinfo {volume} {109}},\ \bibinfo
  {pages} {022002} (\bibinfo {year} {2024}{\natexlab{b}})}\BibitemShut
  {NoStop}%
\bibitem [{\citenamefont {Glaser}\ \emph {et~al.}(2016)\citenamefont {Glaser},
  \citenamefont {Erdmann}, \citenamefont {H\"orandel}, \citenamefont {Huege},\
  and\ \citenamefont {Schulz}}]{Glaser:2016qso}%
  \BibitemOpen
  \bibfield  {author} {\bibinfo {author} {\bibfnamefont {C.}~\bibnamefont
  {Glaser}}, \bibinfo {author} {\bibfnamefont {M.}~\bibnamefont {Erdmann}},
  \bibinfo {author} {\bibfnamefont {J.~R.}\ \bibnamefont {H\"orandel}},
  \bibinfo {author} {\bibfnamefont {T.}~\bibnamefont {Huege}},\ and\ \bibinfo
  {author} {\bibfnamefont {J.}~\bibnamefont {Schulz}},\ }\bibfield  {title}
  {\bibinfo {title} {{Simulation of Radiation Energy Release in Air Showers}},\
  }\href {https://doi.org/10.1088/1475-7516/2016/09/024} {\bibfield  {journal}
  {\bibinfo  {journal} {JCAP}\ }\textbf {\bibinfo {volume} {09}},\ \bibinfo
  {pages} {024}},\ \Eprint {https://arxiv.org/abs/1606.01641} {arXiv:1606.01641
  [astro-ph.HE]} \BibitemShut {NoStop}%
\bibitem [{\citenamefont {{Kahn}}\ and\ \citenamefont
  {{Lerche}}(1966)}]{1966RSPSA.289..206K}%
  \BibitemOpen
  \bibfield  {author} {\bibinfo {author} {\bibfnamefont {F.~D.}\ \bibnamefont
  {{Kahn}}}\ and\ \bibinfo {author} {\bibfnamefont {I.}~\bibnamefont
  {{Lerche}}},\ }\bibfield  {title} {\bibinfo {title} {{Radiation from Cosmic
  Ray Air Showers}},\ }\href {https://doi.org/10.1098/rspa.1966.0007}
  {\bibfield  {journal} {\bibinfo  {journal} {Proceedings of the Royal Society
  of London Series A}\ }\textbf {\bibinfo {volume} {289}},\ \bibinfo {pages}
  {206} (\bibinfo {year} {1966})}\BibitemShut {NoStop}%
\bibitem [{\citenamefont {Ardouin}\ \emph {et~al.}(2005)\citenamefont {Ardouin}
  \emph {et~al.}}]{Ardouin:2005}%
  \BibitemOpen
  \bibfield  {author} {\bibinfo {author} {\bibfnamefont {D.}~\bibnamefont
  {Ardouin}} \emph {et~al.} (\bibinfo {collaboration} {CODALEMA}),\ }\bibfield
  {title} {\bibinfo {title} {{Radio-detection signature of high-energy cosmic
  rays by the CODALEMA experiment}},\ }\href
  {https://doi.org/https://doi.org/10.1016/j.nima.2005.08.096} {\bibfield
  {journal} {\bibinfo  {journal} {Nuclear Instruments and Methods in Physics
  Research Section A: Accelerators, Spectrometers, Detectors and Associated
  Equipment}\ }\textbf {\bibinfo {volume} {555}},\ \bibinfo {pages} {148}
  (\bibinfo {year} {2005})}\BibitemShut {NoStop}%
\bibitem [{\citenamefont {Falcke}\ \emph {et~al.}(2005)\citenamefont {Falcke}
  \emph {et~al.}}]{Falcke:2005}%
  \BibitemOpen
  \bibfield  {author} {\bibinfo {author} {\bibfnamefont {H.}~\bibnamefont
  {Falcke}} \emph {et~al.} (\bibinfo {collaboration} {LOPES}),\ }\bibfield
  {title} {\bibinfo {title} {Detection and imaging of atmospheric radio flashes
  from cosmic ray air showers},\ }\href
  {https://doi.org/https://doi.org/10.1038/nature03614} {\bibfield  {journal}
  {\bibinfo  {journal} {Nature}\ }\textbf {\bibinfo {volume} {435}},\ \bibinfo
  {pages} {313} (\bibinfo {year} {2005})}\BibitemShut {NoStop}%
\bibitem [{\citenamefont {Huege}\ \emph {et~al.}(2019)\citenamefont {Huege}
  \emph {et~al.}}]{Huege:2019}%
  \BibitemOpen
  \bibfield  {author} {\bibinfo {author} {\bibfnamefont {T.}~\bibnamefont
  {Huege}} \emph {et~al.} (\bibinfo {collaboration} {Pierre Auger}),\
  }\bibfield  {title} {\bibinfo {title} {{Radio detection of cosmic rays with
  the Auger Engineering Radio Array}},\ }\href
  {https://doi.org/10.1051/epjconf/201921005011} {\bibfield  {journal}
  {\bibinfo  {journal} {EPJ Web Conf.}\ }\textbf {\bibinfo {volume} {210}},\
  \bibinfo {pages} {05011} (\bibinfo {year} {2019})}\BibitemShut {NoStop}%
\bibitem [{\citenamefont {{van Haarlem}}\ \emph {et~al.}(2013)\citenamefont
  {{van Haarlem}} \emph {et~al.}}]{vanHaarlem:2013}%
  \BibitemOpen
  \bibfield  {author} {\bibinfo {author} {\bibfnamefont {M.~P.}\ \bibnamefont
  {{van Haarlem}}} \emph {et~al.},\ }\bibfield  {title} {\bibinfo {title}
  {{LOFAR: The Low Frequency Array}},\ }\href@noop {} {\bibfield  {journal}
  {\bibinfo  {journal} {Astronomy and Astrophysics}\ }\textbf {\bibinfo
  {volume} {556}},\ \bibinfo {pages} {A2} (\bibinfo {year} {2013})}\BibitemShut
  {NoStop}%
\bibitem [{\citenamefont {Corstanje}\ \emph {et~al.}(2015)\citenamefont
  {Corstanje} \emph {et~al.}}]{Corstanje:2014waa}%
  \BibitemOpen
  \bibfield  {author} {\bibinfo {author} {\bibfnamefont {A.}~\bibnamefont
  {Corstanje}} \emph {et~al.} (\bibinfo {collaboration} {LOFAR}),\ }\bibfield
  {title} {\bibinfo {title} {{The shape of the radio wavefront of extensive air
  showers as measured with LOFAR}},\ }\href
  {https://doi.org/10.1016/j.astropartphys.2014.06.001} {\bibfield  {journal}
  {\bibinfo  {journal} {Astropart. Phys.}\ }\textbf {\bibinfo {volume} {61}},\
  \bibinfo {pages} {22} (\bibinfo {year} {2015})},\ \Eprint
  {https://arxiv.org/abs/1404.3907} {arXiv:1404.3907 [astro-ph.HE]}
  \BibitemShut {NoStop}%
\bibitem [{\citenamefont {Schellart}\ \emph {et~al.}(2014)\citenamefont
  {Schellart} \emph {et~al.}}]{Schellart:2014oaa}%
  \BibitemOpen
  \bibfield  {author} {\bibinfo {author} {\bibfnamefont {P.}~\bibnamefont
  {Schellart}} \emph {et~al.} (\bibinfo {collaboration} {LOFAR}),\ }\bibfield
  {title} {\bibinfo {title} {{Polarized radio emission from extensive air
  showers measured with LOFAR}},\ }\href
  {https://doi.org/10.1088/1475-7516/2014/10/014} {\bibfield  {journal}
  {\bibinfo  {journal} {JCAP}\ }\textbf {\bibinfo {volume} {10}},\ \bibinfo
  {pages} {014}},\ \Eprint {https://arxiv.org/abs/1406.1355} {arXiv:1406.1355
  [astro-ph.HE]} \BibitemShut {NoStop}%
\bibitem [{\citenamefont {Nelles}\ \emph {et~al.}(2015)\citenamefont {Nelles}
  \emph {et~al.}}]{Nelles:2014dja}%
  \BibitemOpen
  \bibfield  {author} {\bibinfo {author} {\bibfnamefont {A.}~\bibnamefont
  {Nelles}} \emph {et~al.} (\bibinfo {collaboration} {LOFAR}),\ }\bibfield
  {title} {\bibinfo {title} {{Measuring a Cherenkov ring in the radio emission
  from air showers at 110\textendash{}190 MHz with LOFAR}},\ }\href
  {https://doi.org/10.1016/j.astropartphys.2014.11.006} {\bibfield  {journal}
  {\bibinfo  {journal} {Astropart. Phys.}\ }\textbf {\bibinfo {volume} {65}},\
  \bibinfo {pages} {11} (\bibinfo {year} {2015})},\ \Eprint
  {https://arxiv.org/abs/1411.6865} {arXiv:1411.6865 [astro-ph.IM]}
  \BibitemShut {NoStop}%
\bibitem [{\citenamefont {Scholten}\ \emph {et~al.}(2016)\citenamefont
  {Scholten} \emph {et~al.}}]{Scholten:2016gmj}%
  \BibitemOpen
  \bibfield  {author} {\bibinfo {author} {\bibfnamefont {O.}~\bibnamefont
  {Scholten}} \emph {et~al.} (\bibinfo {collaboration} {LOFAR}),\ }\bibfield
  {title} {\bibinfo {title} {{Measurement of the circular polarization in radio
  emission from extensive air showers confirms emission mechanisms}},\ }\href
  {https://doi.org/10.1103/PhysRevD.94.103010} {\bibfield  {journal} {\bibinfo
  {journal} {Phys. Rev. D}\ }\textbf {\bibinfo {volume} {94}},\ \bibinfo
  {pages} {103010} (\bibinfo {year} {2016})},\ \Eprint
  {https://arxiv.org/abs/1611.00758} {arXiv:1611.00758 [astro-ph.IM]}
  \BibitemShut {NoStop}%
\bibitem [{\citenamefont {Corstanje}\ \emph {et~al.}(2021)\citenamefont
  {Corstanje} \emph {et~al.}}]{Corstanje:2021kik}%
  \BibitemOpen
  \bibfield  {author} {\bibinfo {author} {\bibfnamefont {A.}~\bibnamefont
  {Corstanje}} \emph {et~al.} (\bibinfo {collaboration} {LOFAR}),\ }\bibfield
  {title} {\bibinfo {title} {{Depth of shower maximum and mass composition of
  cosmic rays from 50 PeV to 2 EeV measured with the LOFAR radio telescope}},\
  }\href {https://doi.org/10.1103/PhysRevD.103.102006} {\bibfield  {journal}
  {\bibinfo  {journal} {Phys. Rev. D}\ }\textbf {\bibinfo {volume} {103}},\
  \bibinfo {pages} {102006} (\bibinfo {year} {2021})},\ \Eprint
  {https://arxiv.org/abs/2103.12549} {arXiv:2103.12549 [astro-ph.HE]}
  \BibitemShut {NoStop}%
\bibitem [{\citenamefont {Buitink}\ \emph {et~al.}(2016)\citenamefont {Buitink}
  \emph {et~al.}}]{Buitink:2016nkf}%
  \BibitemOpen
  \bibfield  {author} {\bibinfo {author} {\bibfnamefont {S.}~\bibnamefont
  {Buitink}} \emph {et~al.} (\bibinfo {collaboration} {LOFAR}),\ }\bibfield
  {title} {\bibinfo {title} {{A large light-mass component of cosmic rays at
  $10^{17}$ - $10^{17.5}$ eV from radio observations}},\ }\href
  {https://doi.org/10.1038/nature16976} {\bibfield  {journal} {\bibinfo
  {journal} {Nature}\ }\textbf {\bibinfo {volume} {531}},\ \bibinfo {pages}
  {70} (\bibinfo {year} {2016})},\ \Eprint {https://arxiv.org/abs/1603.01594}
  {arXiv:1603.01594 [astro-ph.HE]} \BibitemShut {NoStop}%
\bibitem [{\citenamefont {Welling}\ \emph {et~al.}(2019)\citenamefont
  {Welling}, \citenamefont {Glaser},\ and\ \citenamefont
  {Nelles}}]{Welling:2019scz}%
  \BibitemOpen
  \bibfield  {author} {\bibinfo {author} {\bibfnamefont {C.}~\bibnamefont
  {Welling}}, \bibinfo {author} {\bibfnamefont {C.}~\bibnamefont {Glaser}},\
  and\ \bibinfo {author} {\bibfnamefont {A.}~\bibnamefont {Nelles}},\
  }\bibfield  {title} {\bibinfo {title} {{Reconstructing the cosmic-ray energy
  from the radio signal measured in one single station}},\ }\href
  {https://doi.org/10.1088/1475-7516/2019/10/075} {\bibfield  {journal}
  {\bibinfo  {journal} {JCAP}\ }\textbf {\bibinfo {volume} {10}},\ \bibinfo
  {pages} {075}},\ \Eprint {https://arxiv.org/abs/1905.11185} {arXiv:1905.11185
  [astro-ph.IM]} \BibitemShut {NoStop}%
\bibitem [{\citenamefont {Razavi-Ghods}\ \emph {et~al.}(2011)\citenamefont
  {Razavi-Ghods}, \citenamefont {Acedo}, \citenamefont {El-Makadema},\ and\
  \citenamefont {Alexander}}]{Razavi:2011}%
  \BibitemOpen
  \bibfield  {author} {\bibinfo {author} {\bibfnamefont {N.}~\bibnamefont
  {Razavi-Ghods}}, \bibinfo {author} {\bibfnamefont {E.}~\bibnamefont {Acedo}},
  \bibinfo {author} {\bibfnamefont {A.}~\bibnamefont {El-Makadema}},\ and\
  \bibinfo {author} {\bibfnamefont {P.}~\bibnamefont {Alexander}},\ }\bibfield
  {title} {\bibinfo {title} {{Analysis of sky contributions to system
  temperature for low frequency SKA aperture array geometries}},\ }\href
  {https://doi.org/10.1007/s10686-011-9278-6} {\bibfield  {journal} {\bibinfo
  {journal} {Experimental Astronomy}\ }\textbf {\bibinfo {volume} {33}}
  (\bibinfo {year} {2011})}\BibitemShut {NoStop}%
\bibitem [{\citenamefont {Navas}\ \emph {et~al.}(2024)\citenamefont {Navas}
  \emph {et~al.}}]{ParticleDataGroup:2024cfk}%
  \BibitemOpen
  \bibfield  {author} {\bibinfo {author} {\bibfnamefont {S.}~\bibnamefont
  {Navas}} \emph {et~al.} (\bibinfo {collaboration} {Particle Data Group}),\
  }\bibfield  {title} {\bibinfo {title} {{Review of particle physics}},\ }\href
  {https://doi.org/10.1103/PhysRevD.110.030001} {\bibfield  {journal} {\bibinfo
   {journal} {Phys. Rev. D}\ }\textbf {\bibinfo {volume} {110}},\ \bibinfo
  {pages} {030001} (\bibinfo {year} {2024})}\BibitemShut {NoStop}%
\bibitem [{\citenamefont {{Thoudam, S.}}\ \emph {et~al.}(2016)\citenamefont
  {{Thoudam, S.}} \emph {et~al.}}]{Thoudam:2016}%
  \BibitemOpen
  \bibfield  {author} {\bibinfo {author} {\bibnamefont {{Thoudam, S.}}} \emph
  {et~al.},\ }\bibfield  {title} {\bibinfo {title} {Cosmic-ray energy spectrum
  and composition up to the ankle: the case for a second {Galactic}
  component},\ }\href {https://doi.org/10.1051/0004-6361/201628894} {\bibfield
  {journal} {\bibinfo  {journal} {A\&A}\ }\textbf {\bibinfo {volume} {595}},\
  \bibinfo {pages} {A33} (\bibinfo {year} {2016})}\BibitemShut {NoStop}%
\bibitem [{\citenamefont {Brancus}\ \emph {et~al.}(2005)\citenamefont {Brancus}
  \emph {et~al.}}]{Brancus:2005ht}%
  \BibitemOpen
  \bibfield  {author} {\bibinfo {author} {\bibfnamefont {I.~M.}\ \bibnamefont
  {Brancus}} \emph {et~al.},\ }\bibfield  {title} {\bibinfo {title} {{The
  cosmic ray experiment KASCADE-Grande}},\ }in\ \href
  {https://doi.org/10.1142/9789812772862_0028} {\emph {\bibinfo {booktitle}
  {{Carpathian Summer School of Physics 2005: Exotic Nuclei and Nuclear /
  Particle Astrophysics}}}}\ (\bibinfo {year} {2005})\ pp.\ \bibinfo {pages}
  {244--252}\BibitemShut {NoStop}%
\bibitem [{\citenamefont {Abbasi}\ \emph
  {et~al.}(2021{\natexlab{a}})\citenamefont {Abbasi} \emph
  {et~al.}}]{Abbasi:2021RF}%
  \BibitemOpen
  \bibfield  {author} {\bibinfo {author} {\bibfnamefont {R.}~\bibnamefont
  {Abbasi}} \emph {et~al.} (\bibinfo {collaboration} {IceCube}),\ }\bibfield
  {title} {\bibinfo {title} {{Study of mass composition of cosmic rays with
  IceTop and IceCube}},\ }\href {https://doi.org/10.22323/1.395.0323}
  {\bibfield  {journal} {\bibinfo  {journal} {PoS}\ }\textbf {\bibinfo {volume}
  {ICRC2021}},\ \bibinfo {pages} {323} (\bibinfo {year}
  {2021}{\natexlab{a}})}\BibitemShut {NoStop}%
\bibitem [{\citenamefont {Aab}\ \emph {et~al.}(2014)\citenamefont {Aab} \emph
  {et~al.}}]{Auger_depth:2014}%
  \BibitemOpen
  \bibfield  {author} {\bibinfo {author} {\bibfnamefont {A.}~\bibnamefont
  {Aab}} \emph {et~al.} (\bibinfo {collaboration} {Pierre Auger}),\ }\bibfield
  {title} {\bibinfo {title} {Depth of maximum of air-shower profiles at the
  {Pierre Auger Observatory. I. Measurements} at energies above
  {$1{0}^{17.8}\text{ }\mathrm{eV}$}},\ }\href
  {https://doi.org/10.1103/PhysRevD.90.122005} {\bibfield  {journal} {\bibinfo
  {journal} {Phys. Rev. D}\ }\textbf {\bibinfo {volume} {90}},\ \bibinfo
  {pages} {122005} (\bibinfo {year} {2014})}\BibitemShut {NoStop}%
\bibitem [{\citenamefont {Abbasi}\ \emph
  {et~al.}(2021{\natexlab{b}})\citenamefont {Abbasi} \emph
  {et~al.}}]{TALE:2020}%
  \BibitemOpen
  \bibfield  {author} {\bibinfo {author} {\bibfnamefont {R.~U.}\ \bibnamefont
  {Abbasi}} \emph {et~al.} (\bibinfo {collaboration} {Telescope Array}),\
  }\bibfield  {title} {\bibinfo {title} {{The Cosmic-Ray Composition between 2
  PeV and 2 EeV Observed with the TALE Detector in Monocular Mode}},\ }\href
  {https://doi.org/10.3847/1538-4357/abdd30} {\bibfield  {journal} {\bibinfo
  {journal} {Astrophys. J.}\ }\textbf {\bibinfo {volume} {909}},\ \bibinfo
  {pages} {178} (\bibinfo {year} {2021}{\natexlab{b}})},\ \Eprint
  {https://arxiv.org/abs/2012.10372} {arXiv:2012.10372 [astro-ph.HE]}
  \BibitemShut {NoStop}%
\bibitem [{\citenamefont {Knurenko}\ \emph {et~al.}(2015)\citenamefont
  {Knurenko} \emph {et~al.}}]{Knurenko:2015}%
  \BibitemOpen
  \bibfield  {author} {\bibinfo {author} {\bibfnamefont {S.}~\bibnamefont
  {Knurenko}} \emph {et~al.},\ }\bibfield  {title} {\bibinfo {title} {Mass
  composition of cosmic rays in the energy region {$10^{16}$ - $10^{18}$ eV by
  data the Small Cherenkov Array at Yakutsk. Comparison with other Arrays}},\
  }\href@noop {} {\bibfield  {journal} {\bibinfo  {journal} {Proceedings of the
  34th International Cosmic Ray Conference, The Hague, The Netherlands, PoS}\
  ,\ \bibinfo {pages} {254}} (\bibinfo {year} {2015})}\BibitemShut {NoStop}%
\bibitem [{\citenamefont {Ostapchenko}(2013)}]{Ostapchenko:2013}%
  \BibitemOpen
  \bibfield  {author} {\bibinfo {author} {\bibfnamefont {S.}~\bibnamefont
  {Ostapchenko}},\ }\bibfield  {title} {\bibinfo {title} {{QGSJET-II:} physics,
  recent improvements, and results for air showers},\ }\href
  {https://doi.org/10.1051/epjconf/20125202001} {\bibfield  {journal} {\bibinfo
   {journal} {EPJ Web of Conferences}\ }\textbf {\bibinfo {volume} {52}},\
  \bibinfo {pages} {02001} (\bibinfo {year} {2013})}\BibitemShut {NoStop}%
\bibitem [{\citenamefont {Pierog}\ \emph {et~al.}(2015)\citenamefont {Pierog},
  \citenamefont {Karpenko}, \citenamefont {Katzy}, \citenamefont {Yatsenko},\
  and\ \citenamefont {Werner}}]{EPOSLHC:2013}%
  \BibitemOpen
  \bibfield  {author} {\bibinfo {author} {\bibfnamefont {T.}~\bibnamefont
  {Pierog}}, \bibinfo {author} {\bibfnamefont {I.}~\bibnamefont {Karpenko}},
  \bibinfo {author} {\bibfnamefont {J.~M.}\ \bibnamefont {Katzy}}, \bibinfo
  {author} {\bibfnamefont {E.}~\bibnamefont {Yatsenko}},\ and\ \bibinfo
  {author} {\bibfnamefont {K.}~\bibnamefont {Werner}},\ }\bibfield  {title}
  {\bibinfo {title} {{EPOS LHC:} test of collective hadronization with data
  measured at the {CERN Large Hadron Collider}},\ }\href
  {https://doi.org/10.1103/PhysRevC.92.034906} {\bibfield  {journal} {\bibinfo
  {journal} {Phys. Rev. C}\ }\textbf {\bibinfo {volume} {92}},\ \bibinfo
  {pages} {034906} (\bibinfo {year} {2015})}\BibitemShut {NoStop}%
\bibitem [{\citenamefont {Riehn}\ \emph
  {et~al.}(2020{\natexlab{a}})\citenamefont {Riehn}, \citenamefont {Engel},
  \citenamefont {Fedynitch}, \citenamefont {Gaisser},\ and\ \citenamefont
  {Stanev}}]{Sibyll:2020}%
  \BibitemOpen
  \bibfield  {author} {\bibinfo {author} {\bibfnamefont {F.}~\bibnamefont
  {Riehn}}, \bibinfo {author} {\bibfnamefont {R.}~\bibnamefont {Engel}},
  \bibinfo {author} {\bibfnamefont {A.}~\bibnamefont {Fedynitch}}, \bibinfo
  {author} {\bibfnamefont {T.~K.}\ \bibnamefont {Gaisser}},\ and\ \bibinfo
  {author} {\bibfnamefont {T.}~\bibnamefont {Stanev}},\ }\bibfield  {title}
  {\bibinfo {title} {Hadronic interaction model sibyll 2.3d and extensive air
  showers},\ }\href {https://doi.org/10.1103/PhysRevD.102.063002} {\bibfield
  {journal} {\bibinfo  {journal} {Phys. Rev. D}\ }\textbf {\bibinfo {volume}
  {102}},\ \bibinfo {pages} {063002} (\bibinfo {year}
  {2020}{\natexlab{a}})}\BibitemShut {NoStop}%
\bibitem [{\citenamefont {Buitink}\ \emph
  {et~al.}(2023{\natexlab{a}})\citenamefont {Buitink} \emph
  {et~al.}}]{Buitink:2023reh}%
  \BibitemOpen
  \bibfield  {author} {\bibinfo {author} {\bibfnamefont {S.}~\bibnamefont
  {Buitink}} \emph {et~al.},\ }\bibfield  {title} {\bibinfo {title}
  {{Constraining the cosmic-ray mass composition by measuring the shower length
  with SKA}},\ }\href {https://doi.org/10.22323/1.424.0046} {\bibfield
  {journal} {\bibinfo  {journal} {PoS}\ }\textbf {\bibinfo {volume}
  {ARENA2022}},\ \bibinfo {pages} {046} (\bibinfo {year}
  {2023}{\natexlab{a}})},\ \Eprint {https://arxiv.org/abs/2307.02907}
  {arXiv:2307.02907 [astro-ph.HE]} \BibitemShut {NoStop}%
\bibitem [{\citenamefont {Desmet}\ \emph {et~al.}(2025)\citenamefont {Desmet},
  \citenamefont {Watanabe}, \citenamefont {Huege},\ and\ \citenamefont
  {Buitink}}]{Desmet:2025}%
  \BibitemOpen
  \bibfield  {author} {\bibinfo {author} {\bibfnamefont {M.}~\bibnamefont
  {Desmet}}, \bibinfo {author} {\bibfnamefont {K.}~\bibnamefont {Watanabe}},
  \bibinfo {author} {\bibfnamefont {T.}~\bibnamefont {Huege}},\ and\ \bibinfo
  {author} {\bibfnamefont {S.}~\bibnamefont {Buitink}},\ }\href
  {https://arxiv.org/abs/2505.10459} {\bibinfo {title} {Smiet: Fast and
  accurate synthesis of radio pulses from extensive air shower using simulated
  templates}} (\bibinfo {year} {2025}),\ \Eprint
  {https://arxiv.org/abs/2505.10459} {arXiv:2505.10459 [astro-ph.HE]}
  \BibitemShut {NoStop}%
\bibitem [{\citenamefont {Corstanje}\ \emph {et~al.}(2023)\citenamefont
  {Corstanje} \emph {et~al.}}]{Corstanje:2023uyg}%
  \BibitemOpen
  \bibfield  {author} {\bibinfo {author} {\bibfnamefont {A.}~\bibnamefont
  {Corstanje}} \emph {et~al.},\ }\bibfield  {title} {\bibinfo {title}
  {{Prospects for measuring the longitudinal particle distribution of
  cosmic-ray air showers with SKA}},\ }\href
  {https://doi.org/10.22323/1.424.0024} {\bibfield  {journal} {\bibinfo
  {journal} {PoS}\ }\textbf {\bibinfo {volume} {ARENA2022}},\ \bibinfo {pages}
  {024} (\bibinfo {year} {2023})},\ \Eprint {https://arxiv.org/abs/2303.09249}
  {arXiv:2303.09249 [astro-ph.HE]} \BibitemShut {NoStop}%
\bibitem [{\citenamefont {Heck}\ \emph {et~al.}(1998)\citenamefont {Heck},
  \citenamefont {Knapp}, \citenamefont {Capdevielle}, \citenamefont {Schatz},\
  and\ \citenamefont {Thouw}}]{Heck:1998vt}%
  \BibitemOpen
  \bibfield  {author} {\bibinfo {author} {\bibfnamefont {D.}~\bibnamefont
  {Heck}}, \bibinfo {author} {\bibfnamefont {J.}~\bibnamefont {Knapp}},
  \bibinfo {author} {\bibfnamefont {J.~N.}\ \bibnamefont {Capdevielle}},
  \bibinfo {author} {\bibfnamefont {G.}~\bibnamefont {Schatz}},\ and\ \bibinfo
  {author} {\bibfnamefont {T.}~\bibnamefont {Thouw}},\ }\href@noop {} {\emph
  {\bibinfo {title} {{CORSIKA: A Monte Carlo code to simulate extensive air
  showers}}}},\ \bibinfo {type} {Tech. Rep.}\ (\bibinfo {year}
  {1998})\BibitemShut {NoStop}%
\bibitem [{\citenamefont {Huege}\ \emph {et~al.}(2013)\citenamefont {Huege},
  \citenamefont {Ludwig},\ and\ \citenamefont {James}}]{Huege:2013vt}%
  \BibitemOpen
  \bibfield  {author} {\bibinfo {author} {\bibfnamefont {T.}~\bibnamefont
  {Huege}}, \bibinfo {author} {\bibfnamefont {M.}~\bibnamefont {Ludwig}},\ and\
  \bibinfo {author} {\bibfnamefont {C.~W.}\ \bibnamefont {James}},\ }\bibfield
  {title} {\bibinfo {title} {{Simulating radio emission from air showers with
  CoREAS}},\ }\href {https://doi.org/10.1063/1.4807534} {\bibfield  {journal}
  {\bibinfo  {journal} {AIP Conf. Proc.}\ }\textbf {\bibinfo {volume} {1535}},\
  \bibinfo {pages} {128} (\bibinfo {year} {2013})},\ \Eprint
  {https://arxiv.org/abs/1301.2132} {arXiv:1301.2132 [astro-ph.HE]}
  \BibitemShut {NoStop}%
\bibitem [{\citenamefont {Gottowik}\ \emph {et~al.}(2018)\citenamefont
  {Gottowik}, \citenamefont {Glaser}, \citenamefont {Huege},\ and\
  \citenamefont {Rautenberg}}]{Gottowik:2017wio}%
  \BibitemOpen
  \bibfield  {author} {\bibinfo {author} {\bibfnamefont {M.}~\bibnamefont
  {Gottowik}}, \bibinfo {author} {\bibfnamefont {C.}~\bibnamefont {Glaser}},
  \bibinfo {author} {\bibfnamefont {T.}~\bibnamefont {Huege}},\ and\ \bibinfo
  {author} {\bibfnamefont {J.}~\bibnamefont {Rautenberg}},\ }\bibfield  {title}
  {\bibinfo {title} {{Determination of the absolute energy scale of extensive
  air showers via radio emission: systematic uncertainty of underlying
  first-principle calculations}},\ }\href
  {https://doi.org/10.1016/j.astropartphys.2018.07.004} {\bibfield  {journal}
  {\bibinfo  {journal} {Astropart. Phys.}\ }\textbf {\bibinfo {volume} {103}},\
  \bibinfo {pages} {87} (\bibinfo {year} {2018})},\ \Eprint
  {https://arxiv.org/abs/1712.07442} {arXiv:1712.07442 [astro-ph.HE]}
  \BibitemShut {NoStop}%
\bibitem [{\citenamefont {{Corstanje}}\ \emph {et~al.}(2023)\citenamefont
  {{Corstanje}} \emph {et~al.}}]{2023JInst..18P9005C}%
  \BibitemOpen
  \bibfield  {author} {\bibinfo {author} {\bibfnamefont {A.}~\bibnamefont
  {{Corstanje}}} \emph {et~al.},\ }\bibfield  {title} {\bibinfo {title} {{A
  high-precision interpolation method for pulsed radio signals from cosmic-ray
  air showers}},\ }\href {https://doi.org/10.1088/1748-0221/18/09/P09005}
  {\bibfield  {journal} {\bibinfo  {journal} {Journal of Instrumentation}\
  }\textbf {\bibinfo {volume} {18}}\bibfield  {number} {\bibinfo  {number} {
  (9)},\ \bibinfo {eid} {P09005}},\ }\Eprint {https://arxiv.org/abs/2306.13514}
  {arXiv:2306.13514 [astro-ph.IM]} \BibitemShut {NoStop}%
\bibitem [{NuR()}]{NuRadioMC_code}%
  \BibitemOpen
  \bibinfo {note}
  {\url{https://github.com/nu-radio/NuRadioMC/tree/develop/NuRadioReco/detector}}\BibitemShut
  {NoStop}%
\bibitem [{\citenamefont {Glaser}\ \emph {et~al.}(2020)\citenamefont {Glaser}
  \emph {et~al.}}]{Glaser:2019cws}%
  \BibitemOpen
  \bibfield  {author} {\bibinfo {author} {\bibfnamefont {C.}~\bibnamefont
  {Glaser}} \emph {et~al.},\ }\bibfield  {title} {\bibinfo {title} {{NuRadioMC:
  Simulating the radio emission of neutrinos from interaction to detector}},\
  }\href {https://doi.org/10.1140/epjc/s10052-020-7612-8} {\bibfield  {journal}
  {\bibinfo  {journal} {Eur. Phys. J. C}\ }\textbf {\bibinfo {volume} {80}},\
  \bibinfo {pages} {77} (\bibinfo {year} {2020})},\ \Eprint
  {https://arxiv.org/abs/1906.01670} {arXiv:1906.01670 [astro-ph.IM]}
  \BibitemShut {NoStop}%
\bibitem [{\citenamefont {Zheng}\ \emph {et~al.}(2017)\citenamefont {Zheng}
  \emph {et~al.}}]{Zheng:2016lul}%
  \BibitemOpen
  \bibfield  {author} {\bibinfo {author} {\bibfnamefont {H.}~\bibnamefont
  {Zheng}} \emph {et~al.},\ }\bibfield  {title} {\bibinfo {title} {{An improved
  model of diffuse galactic radio emission from 10 MHz to 5 THz}},\ }\href
  {https://doi.org/10.1093/mnras/stw2525} {\bibfield  {journal} {\bibinfo
  {journal} {Mon. Not. Roy. Astron. Soc.}\ }\textbf {\bibinfo {volume} {464}},\
  \bibinfo {pages} {3486} (\bibinfo {year} {2017})},\ \Eprint
  {https://arxiv.org/abs/1605.04920} {arXiv:1605.04920 [astro-ph.CO]}
  \BibitemShut {NoStop}%
\bibitem [{\citenamefont {Martinelli}\ \emph {et~al.}(2025)\citenamefont
  {Martinelli}, \citenamefont {Huege}, \citenamefont {Ravignani},\ and\
  \citenamefont {Schoorlemmer}}]{Martinelli:2024bzg}%
  \BibitemOpen
  \bibfield  {author} {\bibinfo {author} {\bibfnamefont {S.}~\bibnamefont
  {Martinelli}}, \bibinfo {author} {\bibfnamefont {T.}~\bibnamefont {Huege}},
  \bibinfo {author} {\bibfnamefont {D.}~\bibnamefont {Ravignani}},\ and\
  \bibinfo {author} {\bibfnamefont {H.}~\bibnamefont {Schoorlemmer}},\
  }\bibfield  {title} {\bibinfo {title} {{Quantifying energy fluence and its
  uncertainty for radio emission from particle cascades in the presence of
  noise}},\ }\href {https://doi.org/10.1016/j.astropartphys.2025.103091}
  {\bibfield  {journal} {\bibinfo  {journal} {Astropart. Phys.}\ }\textbf
  {\bibinfo {volume} {168}},\ \bibinfo {pages} {103091} (\bibinfo {year}
  {2025})},\ \Eprint {https://arxiv.org/abs/2407.18654} {arXiv:2407.18654
  [astro-ph.IM]} \BibitemShut {NoStop}%
\bibitem [{\citenamefont {Scholten}\ \emph {et~al.}(2024)\citenamefont
  {Scholten} \emph {et~al.}}]{Scholten:2024upn}%
  \BibitemOpen
  \bibfield  {author} {\bibinfo {author} {\bibfnamefont {O.}~\bibnamefont
  {Scholten}} \emph {et~al.},\ }\bibfield  {title} {\bibinfo {title} {{Aperture
  correction for beamforming in the radiometric detection of ultrahigh energy
  cosmic rays}},\ }\href {https://doi.org/10.1103/PhysRevD.110.103036}
  {\bibfield  {journal} {\bibinfo  {journal} {Phys. Rev. D}\ }\textbf {\bibinfo
  {volume} {110}},\ \bibinfo {pages} {103036} (\bibinfo {year} {2024})},\
  \Eprint {https://arxiv.org/abs/2411.12324} {arXiv:2411.12324 [astro-ph.IM]}
  \BibitemShut {NoStop}%
\bibitem [{\citenamefont {Schoorlemmer}\ and\ \citenamefont
  {Carvalho}(2021)}]{Schoorlemmer:2020low}%
  \BibitemOpen
  \bibfield  {author} {\bibinfo {author} {\bibfnamefont {H.}~\bibnamefont
  {Schoorlemmer}}\ and\ \bibinfo {author} {\bibfnamefont {W.~R.}\ \bibnamefont
  {Carvalho}},\ }\bibfield  {title} {\bibinfo {title} {{Radio interferometry
  applied to the observation of cosmic-ray induced extensive air showers}},\
  }\href {https://doi.org/10.1140/epjc/s10052-021-09925-9} {\bibfield
  {journal} {\bibinfo  {journal} {Eur. Phys. J. C}\ }\textbf {\bibinfo {volume}
  {81}},\ \bibinfo {pages} {1120} (\bibinfo {year} {2021})},\ \Eprint
  {https://arxiv.org/abs/2006.10348} {arXiv:2006.10348 [astro-ph.HE]}
  \BibitemShut {NoStop}%
\bibitem [{\citenamefont {Apel}\ \emph {et~al.}(2014)\citenamefont {Apel} \emph
  {et~al.}}]{Apel:2014usa}%
  \BibitemOpen
  \bibfield  {author} {\bibinfo {author} {\bibfnamefont {W.~D.}\ \bibnamefont
  {Apel}} \emph {et~al.} (\bibinfo {collaboration} {LOPES}),\ }\bibfield
  {title} {\bibinfo {title} {{The wavefront of the radio signal emitted by
  cosmic ray air showers}},\ }\href
  {https://doi.org/10.1088/1475-7516/2014/09/025} {\bibfield  {journal}
  {\bibinfo  {journal} {JCAP}\ }\textbf {\bibinfo {volume} {09}},\ \bibinfo
  {pages} {025}},\ \Eprint {https://arxiv.org/abs/1404.3283} {arXiv:1404.3283
  [hep-ex]} \BibitemShut {NoStop}%
\bibitem [{\citenamefont {Schl\"uter}\ and\ \citenamefont
  {Huege}(2021)}]{Schluter:2021egm}%
  \BibitemOpen
  \bibfield  {author} {\bibinfo {author} {\bibfnamefont {F.}~\bibnamefont
  {Schl\"uter}}\ and\ \bibinfo {author} {\bibfnamefont {T.}~\bibnamefont
  {Huege}},\ }\bibfield  {title} {\bibinfo {title} {{Expected performance of
  air-shower measurements with the radio-interferometric technique}},\ }\href
  {https://doi.org/10.1088/1748-0221/16/07/P07048} {\bibfield  {journal}
  {\bibinfo  {journal} {JINST}\ }\textbf {\bibinfo {volume} {16}}\bibfield
  {number} {\bibinfo  {number} { (07)},\ \bibinfo {pages} {P07048}},\ }\Eprint
  {https://arxiv.org/abs/2102.13577} {arXiv:2102.13577 [astro-ph.IM]}
  \BibitemShut {NoStop}%
\bibitem [{\citenamefont {Corstanje}\ \emph {et~al.}(2023)\citenamefont
  {Corstanje} \emph {et~al.}}]{Corstanje:2023nlk}%
  \BibitemOpen
  \bibfield  {author} {\bibinfo {author} {\bibfnamefont {A.}~\bibnamefont
  {Corstanje}} \emph {et~al.},\ }\bibfield  {title} {\bibinfo {title}
  {{Simulations of radio detection of cosmic rays with SKA-Low}},\ }\href
  {https://doi.org/10.22323/1.444.0500} {\bibfield  {journal} {\bibinfo
  {journal} {PoS}\ }\textbf {\bibinfo {volume} {ICRC2023}},\ \bibinfo {pages}
  {500} (\bibinfo {year} {2023})}\BibitemShut {NoStop}%
\bibitem [{\citenamefont {Gaisser}\ and\ \citenamefont
  {Hillas}(1977)}]{GaiserHillas}%
  \BibitemOpen
  \bibfield  {author} {\bibinfo {author} {\bibfnamefont {T.~K.}\ \bibnamefont
  {Gaisser}}\ and\ \bibinfo {author} {\bibfnamefont {A.~M.}\ \bibnamefont
  {Hillas}},\ }\bibfield  {title} {\bibinfo {title} {{Reliability of the method
  of constant intensity cuts for reconstructing the average development of
  vertical showers}},\ }\href@noop {} {\bibfield  {journal} {\bibinfo
  {journal} {International Cosmic Ray Conference, 15th, Plovdiv, Bulgaria,
  August 13-26, 1977, Conference Papers}\ ,\ \bibinfo {pages} {A79}} (\bibinfo
  {year} {1977})}\BibitemShut {NoStop}%
\bibitem [{\citenamefont {{Matthews}}\ \emph {et~al.}(2010)\citenamefont
  {{Matthews}}, \citenamefont {{Mesler}}, \citenamefont {{Becker}},
  \citenamefont {{Gold}},\ and\ \citenamefont {{Hague}}}]{Matthews2010}%
  \BibitemOpen
  \bibfield  {author} {\bibinfo {author} {\bibfnamefont {J.~A.~J.}\
  \bibnamefont {{Matthews}}}, \bibinfo {author} {\bibfnamefont
  {R.}~\bibnamefont {{Mesler}}}, \bibinfo {author} {\bibfnamefont {B.~R.}\
  \bibnamefont {{Becker}}}, \bibinfo {author} {\bibfnamefont {M.~S.}\
  \bibnamefont {{Gold}}},\ and\ \bibinfo {author} {\bibfnamefont {J.~D.}\
  \bibnamefont {{Hague}}},\ }\bibfield  {title} {\bibinfo {title} {{A
  parameterization of cosmic ray shower profiles based on shower width}},\
  }\href {https://doi.org/10.1088/0954-3899/37/2/025202} {\bibfield  {journal}
  {\bibinfo  {journal} {Journal of Physics G Nuclear Physics}\ }\textbf
  {\bibinfo {volume} {37}},\ \bibinfo {eid} {025202} (\bibinfo {year}
  {2010})},\ \Eprint {https://arxiv.org/abs/0909.4014} {arXiv:0909.4014
  [astro-ph.IM]} \BibitemShut {NoStop}%
\bibitem [{\citenamefont {Andringa}\ \emph {et~al.}(2011)\citenamefont
  {Andringa}, \citenamefont {Conceicao},\ and\ \citenamefont
  {Pimenta}}]{Andringa:2011zz}%
  \BibitemOpen
  \bibfield  {author} {\bibinfo {author} {\bibfnamefont {S.}~\bibnamefont
  {Andringa}}, \bibinfo {author} {\bibfnamefont {R.}~\bibnamefont
  {Conceicao}},\ and\ \bibinfo {author} {\bibfnamefont {M.}~\bibnamefont
  {Pimenta}},\ }\bibfield  {title} {\bibinfo {title} {{Mass composition and
  cross-section from the shape of cosmic ray shower longitudinal profiles}},\
  }\href {https://doi.org/10.1016/j.astropartphys.2010.10.002} {\bibfield
  {journal} {\bibinfo  {journal} {Astropart. Phys.}\ }\textbf {\bibinfo
  {volume} {34}},\ \bibinfo {pages} {360} (\bibinfo {year} {2011})}\BibitemShut
  {NoStop}%
\bibitem [{\citenamefont {Buitink}\ \emph
  {et~al.}(2023{\natexlab{b}})\citenamefont {Buitink} \emph
  {et~al.}}]{Buitink:2023rso}%
  \BibitemOpen
  \bibfield  {author} {\bibinfo {author} {\bibfnamefont {S.}~\bibnamefont
  {Buitink}} \emph {et~al.},\ }\bibfield  {title} {\bibinfo {title}
  {{High-resolution air shower observations with the Square Kilometer Array}},\
  }\href {https://doi.org/10.22323/1.444.0503} {\bibfield  {journal} {\bibinfo
  {journal} {PoS}\ }\textbf {\bibinfo {volume} {ICRC2023}},\ \bibinfo {pages}
  {503} (\bibinfo {year} {2023}{\natexlab{b}})}\BibitemShut {NoStop}%
\bibitem [{\citenamefont {Mitra}\ \emph {et~al.}(2020)\citenamefont {Mitra}
  \emph {et~al.}}]{Mitra:2020mza}%
  \BibitemOpen
  \bibfield  {author} {\bibinfo {author} {\bibfnamefont {P.}~\bibnamefont
  {Mitra}} \emph {et~al.},\ }\bibfield  {title} {\bibinfo {title}
  {{Reconstructing air shower parameters with LOFAR using event specific GDAS
  atmosphere}},\ }\href {https://doi.org/10.1016/j.astropartphys.2020.102470}
  {\bibfield  {journal} {\bibinfo  {journal} {Astropart. Phys.}\ }\textbf
  {\bibinfo {volume} {123}},\ \bibinfo {pages} {102470} (\bibinfo {year}
  {2020})},\ \Eprint {https://arxiv.org/abs/2006.02228} {arXiv:2006.02228
  [astro-ph.HE]} \BibitemShut {NoStop}%
\bibitem [{SKA()}]{SKA-documents}%
  \BibitemOpen
  \bibinfo {note}
  {\url{https://www.skao.int/en/science-users/122/relevant-documents}}\BibitemShut
  {NoStop}%
\bibitem [{\citenamefont {Riehn}\ \emph
  {et~al.}(2020{\natexlab{b}})\citenamefont {Riehn}, \citenamefont {Engel},
  \citenamefont {Fedynitch}, \citenamefont {Gaisser},\ and\ \citenamefont
  {Stanev}}]{Riehn:2019jet}%
  \BibitemOpen
  \bibfield  {author} {\bibinfo {author} {\bibfnamefont {F.}~\bibnamefont
  {Riehn}}, \bibinfo {author} {\bibfnamefont {R.}~\bibnamefont {Engel}},
  \bibinfo {author} {\bibfnamefont {A.}~\bibnamefont {Fedynitch}}, \bibinfo
  {author} {\bibfnamefont {T.~K.}\ \bibnamefont {Gaisser}},\ and\ \bibinfo
  {author} {\bibfnamefont {T.}~\bibnamefont {Stanev}},\ }\bibfield  {title}
  {\bibinfo {title} {{Hadronic interaction model Sibyll 2.3d and extensive air
  showers}},\ }\href {https://doi.org/10.1103/PhysRevD.102.063002} {\bibfield
  {journal} {\bibinfo  {journal} {Phys. Rev. D}\ }\textbf {\bibinfo {volume}
  {102}},\ \bibinfo {pages} {063002} (\bibinfo {year} {2020}{\natexlab{b}})},\
  \Eprint {https://arxiv.org/abs/1912.03300} {arXiv:1912.03300 [hep-ph]}
  \BibitemShut {NoStop}%
\bibitem [{\citenamefont {Buitink}\ \emph {et~al.}(2024)\citenamefont {Buitink}
  \emph {et~al.}}]{Buitink:2024/j}%
  \BibitemOpen
  \bibfield  {author} {\bibinfo {author} {\bibfnamefont {S.}~\bibnamefont
  {Buitink}} \emph {et~al.} (\bibinfo {collaboration} {SKA SWG HECP}),\
  }\bibfield  {title} {\bibinfo {title} {{Cosmic ray observations with the
  Square Kilometer Array}},\ }\href {https://doi.org/10.22323/1.470.0025}
  {\bibfield  {journal} {\bibinfo  {journal} {PoS}\ }\textbf {\bibinfo {volume}
  {ARENA2024}},\ \bibinfo {pages} {025} (\bibinfo {year} {2024})}\BibitemShut
  {NoStop}%
\bibitem [{\citenamefont {Abdul~Halim}\ \emph {et~al.}(2025)\citenamefont
  {Abdul~Halim} \emph {et~al.}}]{PierreAuger:2024flk}%
  \BibitemOpen
  \bibfield  {author} {\bibinfo {author} {\bibfnamefont {A.}~\bibnamefont
  {Abdul~Halim}} \emph {et~al.} (\bibinfo {collaboration} {Pierre Auger}),\
  }\bibfield  {title} {\bibinfo {title} {{Inference of the Mass Composition of
  Cosmic Rays with Energies from $10^{18.5}$ to $10^{20}$\,eV using the Pierre
  Auger Observatory and Deep Learning}},\ }\href
  {https://doi.org/10.1103/PhysRevLett.134.021001} {\bibfield  {journal}
  {\bibinfo  {journal} {Phys. Rev. Lett.}\ }\textbf {\bibinfo {volume} {134}},\
  \bibinfo {pages} {021001} (\bibinfo {year} {2025})},\ \Eprint
  {https://arxiv.org/abs/2406.06315} {arXiv:2406.06315 [astro-ph.HE]}
  \BibitemShut {NoStop}%
\bibitem [{\citenamefont {Thoudam}\ \emph {et~al.}(2016)\citenamefont {Thoudam}
  \emph {et~al.}}]{Thoudam:2015lba}%
  \BibitemOpen
  \bibfield  {author} {\bibinfo {author} {\bibfnamefont {S.}~\bibnamefont
  {Thoudam}} \emph {et~al.} (\bibinfo {collaboration} {LOFAR}),\ }\bibfield
  {title} {\bibinfo {title} {{Measurement of the cosmic-ray energy spectrum
  above 10$^{16}$ eV with the LOFAR Radboud Air Shower Array}},\ }\href
  {https://doi.org/10.1016/j.astropartphys.2015.06.005} {\bibfield  {journal}
  {\bibinfo  {journal} {Astropart. Phys.}\ }\textbf {\bibinfo {volume} {73}},\
  \bibinfo {pages} {34} (\bibinfo {year} {2016})},\ \Eprint
  {https://arxiv.org/abs/1506.09134} {arXiv:1506.09134 [astro-ph.IM]}
  \BibitemShut {NoStop}%
\end{thebibliography}%

\end{document}